\documentclass[pra,twocolumn,groupedaddress,showpacs]{revtex4-1}

\usepackage{amsmath,amssymb}
\usepackage{amsthm}
\usepackage{graphicx}

\newtheorem{lemma}{Lemma}
\newtheorem{theorem}[lemma]{Theorem}
\newtheorem{coro}[lemma]{Corollary}

\def\cites#1{XX Need reference XX}
\long\def\ca#1\cb{}
\def\bra#1{\langle#1|}
\def\dl{\delta }
\def\dyad#1#2{|#1\rangle\langle#2|}
\def\inpd#1#2{\langle#1|#2\rangle }
\def\ket#1{|#1\rangle }
\def\lm{\lambda }

\def\mted#1#2#3{\langle#1|#2|#3\rangle }
\def\ra{\rightarrow }

\def\HC{{\cal H}}
\def\JC{{\cal J}}
\def\QC{{\cal Q}}
\def\QCL{{\cal Q^{(\lambda)}}}
\def\RC{{\cal R}}
\def\RCL{{\cal R^{(\lambda)}}}

\def\UC{{\cal U}}
\def\VC{{\cal V}}
\def\WC{{\cal W}}

\newcommand{\Tr}{{\rm Tr}}

\long\def\ca#1\cb{}

 \def\outl#1{}  \def\xa{} \def\xb{}  

\ca
 \def\outl#1{\par{\medskip\noindent\hspace*{.5cm}\bf
      \mathversion{bold}#1\mathversion{normal}\smallskip} }
 \long\def\xa#1\xb{}
 
\cb

\ca
 \def\outl#1{\par{\medskip\noindent\hspace*{.5cm}\bf
      \mathversion{bold}#1\mathversion{normal}\smallskip} }
 \def\xa{} \def\xb{}  
\cb

\begin{document}


\title{Efficient implementation of bipartite nonlocal unitary gates using prior entanglement and classical communication}

\author{Li Yu$^1$}
\email{liy@andrew.cmu.edu}

\author{Robert B. Griffiths$^1$}
\email{rgrif@andrew.cmu.edu}

\author{Scott M. Cohen$^{1,2}$}
\email{cohensm@duq.edu}
\affiliation{$^1$Department of Physics, Carnegie-Mellon University,
Pittsburgh, Pennsylvania 15213, U.S.A.\\
$^2$Department of Physics, Duquesne University, Pittsburgh,
Pennsylvania 15282, U.S.A.}

\xa
\begin{abstract}
Any bipartite nonlocal unitary operation can be carried out by teleporting a quantum state from one party to the other, performing the unitary gate locally, and teleporting a state back again. This paper investigates unitaries which can be carried out using less prior entanglement and classical communication than are needed for teleportation. Large families of such unitaries are constructed using (projective) representations of finite groups.  Among the tools employed are: a diagrammatic approach for representing entangled states, a theorem on the necessary absence of information at certain times and locations, and a representation of bipartite unitaries based on a group Fourier transform.
\end{abstract}
\xb

\date{Version of June 19, 2010}

\maketitle



\section{Introduction}
\label{sct1}\xa

\xb\outl{Teleportation, alternatives, literature}\xa

Teleportation \cite{Bennett_tele} makes possible a wide variety of nonlocal
quantum processes provided sufficient prior entanglement and classical
communication are available.  To change the quantum state of two systems $A$
and $B$ separated in space it is only necessary to teleport the $A$ state to
the laboratory where $B$ is located, carry out the desired operation, and
teleport the result back again.  The present paper is concerned with certain
types of operations, nonlocal unitaries, which can be achieved at lower cost,
in particular less prior entanglement than is needed for two-way
teleportation.

Protocols of this sort for qubits were developed by Eisert et al.\
\cite{Eisert} and Reznik et al.\ \cite{ReznikNLU}, and our work builds upon
theirs. See
\cite{Gottesman,Chefles,Chefles2,Song,Huang,Chen,OLeary,Zeng,Berry,
Dang,Zhao,Jang,Groisman_nonMES,Cirac,Dur,Ye} for other deterministic and
probabilistic protocols. We only consider the deterministic case in which the
desired unitary is carried out with probability 1; finding efficient protocols
that allow for some noise is a challenging problem not addressed here.  While
our protocols can be applied to an arbitrary nonlocal unitary, only in special
cases are they more efficient than teleportation.

\xb\outl{Motivations minimizing entanglement cost, understanding types of
information}\xa

There are two separate motivations behind studies of the sort presented here.
First, it seems likely that nonlocal operations will play a significant role
in future quantum computers, especially schemes for distributed computation
\cite{Cirac_dist,Yims,Yims2,Meter}, and achieving them using prior
entanglement is an option that deserves consideration.  Since producing
entanglement is likely to be expensive, there is an obvious advantage to
protocols which use as little of it as possible.  By comparison, classical
communication is usually thought of as cheap, since it does not have to be
protected from decoherence.  However, protocols that minimize its use might
still have an advantage over those that require more.

A second motivation is the desire to better understand the role of different
types of quantum information \cite{RBGtypes} in various information-processing
tasks.  General nonlocal unitaries require physical influences to propagate
from one side to the other, in fact in both directions, and one would like to
understand these in rational terms rather than by invoking quasimagical
``collapses'' and the like.  It turns out that unitary operations place
particularly stringent conditions on the presence or absence of different
types of information at different locations and different times throughout a
protocol, and knowing what these are can provide insight into why certain
gates and measurements need to be employed rather than others. In this respect
our analysis represents an advance over earlier work.

\xb\outl{General form of the nonlocal unitary}\xa

Our protocols are based upon expanding a nonlocal unitary on
$\HC_A\otimes\HC_B$ in the form
\begin{equation}
\label{eqn1} \UC=\sum_{j=1}^N A_j\otimes B_j,
\end{equation}
where the Hilbert spaces $\HC_A$, $\HC_B$ have finite dimensions $d_A,d_B$
respectively, the $A_j$ operators satisfy special conditions, or are of a
particular type, and the expansion coefficients $B_j$, obviously constrained
by the requirement that $\UC$ be unitary, may have additional special
properties.  In the simplest situation of ``controlled'' unitaries, the one
easiest to understand in information-theoretic terms, the $A_j$ form a
projective decomposition of the identity on $\HC_A$, and the $B_j$ are
arbitrary unitaries on $\HC_B$.  A more complex, but also more general case is
unitaries of ``group'' form (or ``group-unitaries''), when the $A_j$ are
unitary operators that form a representation, possibly a projective
representation, of a group $G$, and the sum in \eqref{eqn1} is a sum over
elements in this group.  Here the $B_j$ may also be proportional to unitary
operators forming a (projective) representation of the same group $G$, but
that is not a requirement of our protocol. Actually any bipartite unitary can
be written in this form with some suitable choice of the group $G$, thus the
term ``group-unitaries'' should be regarded as describing the form of
expansion of $\UC$, rather than a property of $\UC$ itself. There is some
relationship between our use of groups for nonlocal unitaries and the protocol
of Klappenecker et al. \cite{Klappenecker} for implementing local unitary
gates.

\xb\outl{Outline of paper}\xa

The remainder of this paper is organized as follows: Sec.~\ref{sct2}
indicates the general strategy for our protocols, shows through a diagrammatic
approach why groups are useful, establishes an information theorem that is
useful when discussing unitaries, and a lower bound on how much entanglement
is needed.

The discussion of particular protocols begins in Sec.~\ref{sct3} with
controlled unitaries that generalize \cite{Eisert} and are easily understood
in information-theoretic terms.  The presentation of the main group-unitary
protocol in Sec.~\ref{sct4} begins with a general definition and a quantum
circuit, followed by a detailed analysis in Sec.~\ref{sct4b} of why the
protocol works.  A procedure related to group Fourier transforms is used in
Sec.~\ref{sct4c} to give a general parametrization of unitaries which can be
realized in this fashion.  A particular case in which the $A$ and $B$ systems
are treated in a symmetrical fashion is the subject of Sec.~\ref{sct4d}, and
in Sec.~\ref{sct4e} it is shown that the controlled unitaries of
Sec.~\ref{sct3} can be rewritten in the group form in an efficient way.

Several specific examples of nonlocal unitaries are presented in
Sec.~\ref{sct5}.  This is followed by a summary of the paper in
Sec.~\ref{sct6}, which notes some issues deserving further study. The
appendices contain proofs and other subsidiary material.

\xb\section{General Comments}
\label{sct2}\xa

\xb\subsection{Overall structure of protocols}\label{sct2a}\xa

\begin{figure*}[ht]
\begin{center}
\includegraphics[scale=1.1]{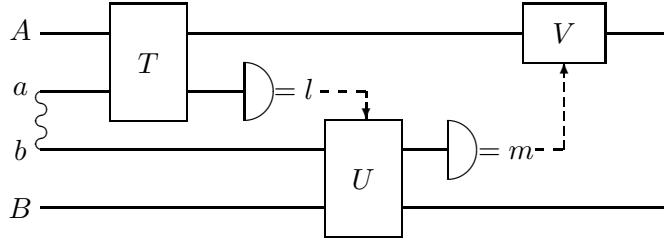}
\end{center}
\caption{A prototypical circuit diagram for local implementation of nonlocal
unitaries} \label{fgr1}
\end{figure*}

The protocols discussed in this paper all have a common structure indicated
schematically in Fig.~\ref{fgr1}. The two parties share an entangled state on
ancillary systems $a$ and $b$. Alice performs a unitary gate $T$ on systems
$a$ and $A$, followed by a measurement of the $a$ system. She sends the
(classical) measurement result to Bob, who uses this to determine which
unitary $U$ to carry out on $b$ and $B$. Bob then measures the $b$ system and
sends the result back to Alice, who uses this information to choose a unitary
$V$ carried out on $A$.  It is worth noting that the straightforward
teleportation protocol mentioned in Sec.~\ref{sct1} can also be represented
using Fig.~\ref{fgr1} provided the initial entangled resource is large enough
to allow teleportation in both directions.  (We leave as an exercise to the
reader working out the details of the unitaries $T$, $U$ and $V$ needed to
achieve this.)  Such protocols are inherently asymmetric in that one party has
to take the first step, while the second party acts according to the
information received from the first party.

\xb\subsection{Why groups are useful}\label{sct2b}\xa

Our most powerful protocols are
based on expansions of the desired nonlocal unitary $\UC$ in terms of a
matrix representation of a finite group. Indeed, all unitaries can be expanded
in this way. Here, we explain why groups are useful utilizing an intuitive and
visual presentation involving diagrams \cite{Cohen_tele,Cohen_UPB}, from which
as an additional benefit we will gain insight into one of our most important
contributions, that unlike in previously published protocols, the Schmidt rank
$N$ of the entangled resource can be chosen independently of the dimensions
$d_A,d_B$ of Hilbert spaces $\HC_A,~\HC_B$ on which $\UC$ acts. We note that
it was by using these diagrams that we discovered the first of our protocols,
which uses a uniformly entangled state to implement unitaries of the form
\begin{equation}\label{eqn2} \UC=\sum_{j=0}^{N-1}c(j)U(j)\otimes V(j),
\end{equation} with $\{U(j)\},\,\{V(j)\}$ two sets of unitaries, and the
operators $U(j) \otimes V(j)$ form an ordinary representation of a
group. All other protocols based on groups were then discovered as
generalizations of this one.

\begin{figure*}[ht]
\begin{center}
\includegraphics[scale=1]{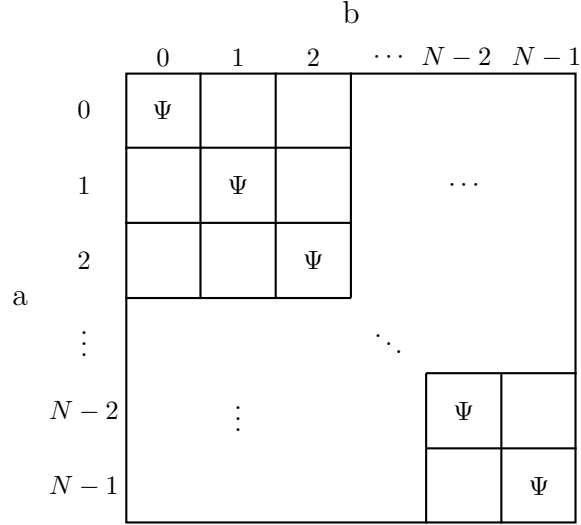}
\end{center}
\caption{Box diagram illustrating how entanglement creates multiple images of
the state $|\Psi\rangle$ distributed along the diagonal of the box.}
\label{fgr2}
\end{figure*}

The idea underlying our diagrams is that the presence of entanglement on
systems $a,b$ effectively creates multiple images of the states of other
systems, in our case, $A$ and $B$ (see \cite{Cohen_tele,Cohen_UPB} for
detailed discussion of these ideas). That is, the state
\begin{equation}\label{eqn3}
|\Phi\rangle\otimes|\Psi\rangle\propto\left(\sum_{j=0}^{N-1}|j\rangle_a|j\rangle_b\right)\otimes|\Psi\rangle,
\end{equation} with $|\Phi\rangle$ on systems $a,b$ and $|\Psi\rangle$ on
$A,B$, can be represented by the diagram shown in Fig.~\ref{fgr2}.

\begin{figure}[ht]
\begin{center}
\includegraphics[scale=1]{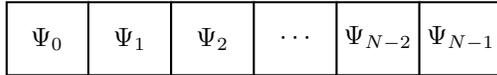}
\end{center}
\caption{Result of projection onto the state $|+\rangle_a$ following their
respective controlled unitaries.}  \label{fgr3}
\end{figure}

Given the way these images are distributed through the diagram,
it is clear that Alice and Bob independently have ``access'' to each of the
individual images, and by performing controlled unitaries,
$\sum_{j=0}^{N-1}|j\rangle_{a}\langle j|\otimes U(j)$ and
$\sum_{j=0}^{N-1}|j\rangle_{b}\langle j|\otimes V(j)$, on each side, can
together attach to each image one of the operators appearing in the terms of
the sum for $\UC$ in \eqref{eqn2}. What we ultimately want is a particular
linear combination of all these terms. To see how to accomplish this, note
that a measurement on $a$ with outcome $j$ picks out a single row of the
diagram, and thus a single image. On the other hand, a measurement outcome
$|+\rangle_a=\sum_{j=0}^{N-1}|j\rangle_a$, places a sum of all the rows into a
single row, which will appear as in Fig.~\ref{fgr3} (all other rows
disappearing), with each column containing one of the terms
$|\Psi_j\rangle=(U(j)\otimes V(j))|\Psi\rangle$. Similarly, if Alice
measures $a$ in the Fourier basis (or equivalently, performs a discrete
Fourier transform on $a$ and then measures in the standard basis), then for
any outcome the diagram is collapsed into a single row, with each column
containing one of the terms $|\Psi_j\rangle$ multiplied by a phase
factor. These phase factors, $e^{i\xi_j}$, can easily be removed by Bob
performing $\sum_{j=0}^{N-1}e^{-i\xi_j}|j\rangle_b\langle j|$ on system $b$
leaving, for all outcomes of Alice's measurement on $a$, the diagram in
Fig.~\ref{fgr3}.

The next step is for Bob to make a measurement of his own on system $b$,
designed to place the factors $c(j)$ on the appropriate images and to take
linear sums of these terms, collapsing the diagram finally into a single small
box. If he first performs a unitary $C$, the first row of which is given by
$\sum_{j=0}^{N-1}c(j)|0\rangle_b\langle j|$, and then measures in the standard
basis with outcome corresponding to state $|0\rangle_b$, the result will be
precisely the desired unitary operation,
\begin{equation}\label{eqn4}
\UC|\Psi\rangle=\sum_{j=0}^{N-1}c(j)|\Psi_j\rangle.
\end{equation} The question is how to complete this unitary, $C$. Clearly, we
cannot simply copy the first row into all the others, as the rows of a unitary
matrix must all be mutually orthogonal, even though this would produce
\eqref{eqn4} for every outcome of Bob's subsequent measurement. One
possibility that turns out will always work (see Sec.~\ref{sct4d}) is to
choose all the other rows of $C$ as permutations of its first row. This choice
does not automatically yield \eqref{eqn4}, however, but we instead obtain
\begin{equation}\label{eqn5}
\UC^\prime|\Psi\rangle=\sum_{j=0}^{N-1}c(\Pi(j))|\Psi_j\rangle,
\end{equation} where $\Pi(j)$ represents the permutation for the given outcome
and $\UC^\prime\ne\UC$, implying that our protocol has failed unless we can
find a way to correct it. Such a correction is certainly possible if there
exist local unitaries on $\HC_A,\HC_B$ that together transform
$|\Psi_j\rangle$ to $|\Psi_{\Pi(j)}\rangle$, and hence $\UC^\prime$ into
$\UC$. This is where the notion of groups enters the picture. Indeed,
recalling that $|\Psi_j\rangle=\left(U(j)\otimes V(j)\right)|\Psi\rangle$,
then if Alice and Bob perform $U(k)\otimes V(k)$, and can choose $k$ such that
$U(k)U(j)=U(\Pi(j))$ and $V(k)V(j)=V(\Pi(j))$ with $\Pi(j)$
corresponding to the multiplication table of a group $G$, then this transforms
$\UC^\prime$ into $\UC$ and we have successfully, and deterministically,
implemented $\UC$. See Sec.~\ref{sct4d} for further details. Note that the
requirement expressed in the previous discussion is that $U,V$ each form a
representation of the group $G$. One could consider the possibility that they
are non-unitary representations, but the corrections just described would
then not generally be possible deterministically. Therefore, we consider only
unitary representations in the sequel.

Finally, we note that it should be clear from this discussion that $N$ can be
chosen as a completely independent quantity; in particular, there is no reason
for it to be constrained by the values of $d_A$ or $d_B$, as has been the case
in previously known protocols. Hence, the amount of entanglement needed to
implement a given nonlocal unitary depends only on the form of the unitary
itself---in particular for our protocols, on the possible ways it can be
expanded in terms of a group---and not on the size of the local Hilbert spaces
upon which it acts.

\xb\subsection{Information location}
\label{sct2c}\xa

\xb\outl{``Quantized'' circuit}\xa

\begin{figure}[ht]
\begin{center}
\includegraphics[scale=1]{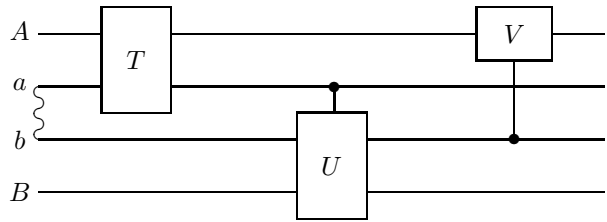}
\end{center}
\caption{Quantized version of Fig.~\ref{fgr1}.}
\label{fgr4}
\end{figure}

To understand various aspects of our protocols, it will be useful to
``quantize'' the circuit in Fig.~\ref{fgr1}, by replacing measurements and
classical communication with appropriate controlled unitaries,
Fig.~\ref{fgr4}.  An easy way of seeing the equivalence of the two circuits in
terms of their action on systems $A$ and $B$ is to imagine that the ancillary
systems in Fig.~\ref{fgr4} are measured in the standard basis at a time
corresponding to the right side of the diagram. Then using the fact that
quantum measurements reveal pre-existing properties when one employs an
appropriate framework \cite{note_framework}, one can infer that systems $a$
and $b$ were in the states indicated by these outcomes, $\ket{l}$ and
$\ket{m}$ respectively, \emph{before} the measurements took place and thus
also at times preceding the controlled operations, i.e., at the times when the
measurements in Fig.~\ref{fgr1} took place.

While Figs.~\ref{fgr1} and \ref{fgr4} are equivalent so far as $A$ and $B$ are
concerned, the latter is simpler to analyze in terms of certain
information-theoretical ideas by using the following theorem.

\begin{theorem}
\label{thm1}

If $\JC$ is an isometry mapping a Hilbert space $\HC_R$ to a tensor product
$\HC_R\otimes\HC_S$, the following statements are equivalent:

{\rm(i)} $\JC$ generates a unitary from $\HC_R$ to itself in the sense that
\begin{equation} \JC \ket{r} = (\UC\ket{r})\otimes \ket{s_0},
\label{eqn6}
\end{equation} with $\UC:\HC_A\ra \HC_A$ a unitary operator and $\ket{s_0}$ a
fixed state (independent of $\ket{r}$) in $\HC_S$.

{\rm(ii)} There is no information about the initial state of $R$ available in
$S$ after applying $\JC$.

{\rm(iii)} The Kraus operators $\{K_j\}$ such that
\begin{equation}
\label{eqn7} \JC \ket{r} = \sum_j K_j \ket{r} \otimes \ket{s_j},
\end{equation} for $\{\ket{s_j}\}$ some orthonormal basis of $\HC_S$, are of
the form $K_j=c_j\WC$, with the $c_j$ complex numbers and $\WC$ a map from
$\HC_R$ to itself.

\end{theorem}

By ``no information'' we mean that the final state of $S$ is uncorrelated with
the initial state of $R$, see \cite{RBG05}, and thus no conceivable
measurement on $S$ will yield any information about the initial state of $R$.
It is then obvious that (i) implies (ii) and (iii); the reverse inferences are
proved in Appendix~\ref{apx1}.

To apply this theorem to the situation in Fig.~\ref{fgr4} let $\VC$ be the
unitary represented by this quantum circuit, $\ket{r}$ the initial state of
the combined $A$ and $B$ system thought of as $R$, $\ket{\Phi}$ the (fixed)
initial state of $a$ and $b$, which together constitute the system $S$, and
\begin{equation} \JC \ket{r} := \VC(\ket{r}\otimes\ket{\Phi}).
\label{eqn8}
\end{equation} The theorem then tells us that if the circuit carries out a
unitary $\UC$ on $\HC_A\otimes\HC_B$ it must be the case that no information
about the initial state of $AB$ can be found in the final state of the
combined ancillary systems $a$ and $b$, and therefore none is present in
either $a$ or $b$ at a time \emph{after} the controlled gate that represents
its final interaction with the rest of the system.  We will see in
Sec.~\ref{sct3} how this absence of information can be used to motivate the
choice of certain parts of a circuit to carry out nonlocal unitaries.  Of
course at a time \emph{before} the final controlled interactions, the
ancillary systems are (in general) correlated with, and thus contain
information about, the rest of the system; it is only after the final
interactions, the choice of which is constrained by the need to remove this
information from the ancillas, that they are in an appropriate sense of the
term ``information-free''.  Since the measurement outcomes in Fig.~\ref{fgr1}
are exactly the same as if the measurements were carried out at the end of the
time interval shown in Fig.~\ref{fgr4}, we see that such outcomes
\emph{cannot} contain any information about the input $AB$.  On the other hand,
there certainly exist measurements that Alice could carry out on $a$ that would
indeed reveal information about the input state of $AB$, and in that case
we cannot implement deterministically a unitary $\UC$ on $\HC_A\otimes\HC_B$.
Conversely, if there is no information about the $AB$ input in the final state
of the ancillaries, then the circuit in Fig.~\ref{fgr4}, and hence its
counterpart in Fig.~\ref{fgr1}, will necessarily result in a unitary map of
the $AB$ input to the $AB$ output.

Statement (iii) of the theorem provides an alternative way of deciding if a circuit
implements a unitary on $R$.  Suppose the whole circuit is (or is equivalent
to) a unitary followed by a measurement in an orthonormal basis of $\HC_S$. If
for every measurement outcome the corresponding Kraus operator on $R$ is
proportional to the same operator, then the circuit deterministically
implements a unitary on $R$. Checking this may be easier than working out
the entire state evolution for an arbitrary initial state of $R$.

\xb\subsection{Bounds on resources}
\label{sct2d}\xa

\xb\outl{Theorem for lower bound of entanglement cost}\xa

The theorem that follows provides a very general lower bound on the amount of
entanglement resource needed to implement a nonlocal unitary.  By the Schmidt
rank (sometimes called ``Schmidt number'') \cite{opSchmidt} of a bipartite
operator on $\HC_A\otimes \HC_B$ we mean the minimum number of terms in an
expansion of the operator as a sum of products of operators on the separate
systems, or, equivalently, the rank of the matrix of coefficients $c_{ij}$
when the operator is expanded in the form $\sum_{ij}c_{ij}A_i\otimes B_j$ with
$\{A_i\}$ and $\{B_j\}$ bases for the operators on $\HC_A$ and $\HC_B$. Either
definition has an obvious counterpart in the Schmidt rank of a bipartite pure
state.
There are various ways of defining the ``entanglement'' of a bipartite unitary operator.  For our purposes a useful
one is the \emph{entangling strength} \cite{Nielsen} of the unitary $\UC_{AB}$
defined as the maximum entanglement of a state on $\HC_{A\bar A}\otimes
\HC_{B\bar B}$ produced by letting $\UC_{AB}\otimes I_{\bar A}\otimes I_{\bar
  B}$ act on a product state $\ket{\sigma}_{A\bar A}\otimes \ket{\tau}_{B\bar
  B}$, where $\bar A$ and $\bar B$ are ancillary systems, and
$\ket{\sigma}_{A\bar A}$ and $\ket{\tau}_{B\bar B}$ are arbitrary entangled
states. By the entanglement of a pure state we mean the usual measure $-\sum
\lambda_j\log \lambda_j$ in terms of its (squared) Schmidt coefficients
$\lambda_j$.

\begin{theorem}
  \label{thm2} To generate a bipartite nonlocal unitary $\UC$ using
  an entangled resource $\ket{\Phi}$ and separable quantum operations
  \cite{Bennett_sep}, which include local operations and classical
  communication (LOCC), the Schmidt rank of $\ket{\Phi}$ cannot be less than the
  Schmidt rank of $\UC$, and the entanglement of $\ket{\Phi}$ cannot be less
  than the entangling strength of $\UC$ as defined previously.
\end{theorem}

The proof will be found in Appendix~\ref{apx2}. The second assertion remains
true if  ``entanglement'' and ``entangling strength'' both refer to some other
entanglement monotone \cite{Vidal_ent}.

The idea behind the second part of Theorem~\ref{thm2} is that entanglement cannot increase on average under LOCC. Using the same idea, the entanglement that a bipartite unitary can generate for any particular input product state provides a lower bound for the entanglement needed to implement this unitary. The entanglement that unitaries of the form \eqref{eqn2} can generate for the input product state $\ket{0}_A\otimes\ket{0}_B$ has effectively been studied in \cite{Hamma}.

\xb\section{Implementing Controlled Unitaries}
\label{sct3}\xa

\xb\outl{Circuit diagram}\xa
\begin{figure*}[ht]
\begin{center}
\includegraphics[scale=1]{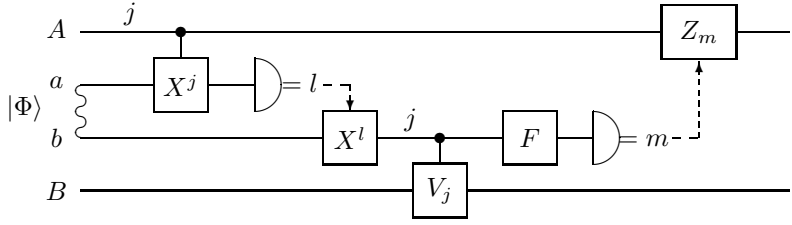}
\end{center}
\caption{$Z$-information protocol when $P_j$ in \eqref{eqn9} projects onto
$\ket{j}_A$. The $j$ label appears twice, reflecting the fact that the $Z$ type of information about system $A$ is transmitted to system $b$.}
\label{fgr5}
\end{figure*}

\begin{figure*}[ht]
\begin{center}
\includegraphics[scale=1]{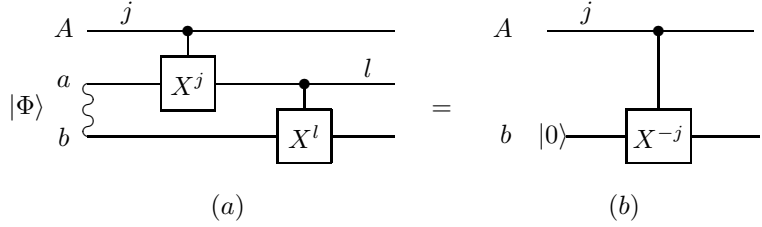}
\end{center}
\caption{Quantized upper left part of Fig.~\ref{fgr5}, and an effectively
equivalent diagram.} \label{fgr6}
\end{figure*}

In this section we consider bipartite controlled unitaries of the form
\begin{equation}\label{eqn9} {\UC}=\sum_{j=0}^{N-1}P_j\otimes V_j,
\end{equation} where the $P_j$'s form a (projective) decomposition of the
identity on $\HC_A$, while the $V_j$'s are arbitrary unitaries on $\HC_B$. All
the essential ideas can be understood assuming the $P_j$'s are of rank 1
(i.e., onto pure states) and this is assumed in the exposition that follows,
with some remarks toward the end about the extension to a more general
situation [see the material associated with \eqref{eqn16}].

\subsection{Description of the protocol with rank-1 projectors $P_j$} \label{sct3a}
Let $\HC_A$ be of dimension $N$ and $P_j$ a projector onto $\ket{j}$ in the
standard basis.  Then the circuit in Fig.~\ref{fgr5} represents a
straightforward generalization of the $N=2$ protocol of \cite{Eisert} to
arbitrary $N$.  The entangled resource is
\begin{equation}\label{eqn10} \vert \Phi\rangle_{ab}
=\frac{1}{\sqrt{N}}\sum_{k=0}^{N-1}\vert k\rangle\otimes\vert k\rangle.
\end{equation}
For general $N$ define the $X$ and $Z$ gates using
\begin{equation}
  \label{eqn11} X\ket{k} = \ket{k-1};\quad Z\ket{k} = e^{2\pi ik/N}\ket{k},
\end{equation}
with subtraction understood as mod $N$, so one has the usual Pauli operators
when $N=2$.  The first controlled-$X^j$ gate in the figure means that if
$\HC_A$ is in the state $\ket{j}$, then $X^j$, meaning $X$ to the power $j$,
is applied to the $a$ system. Then Alice performs a measurement on $a$ in the
standard basis, and sends the result $l$ to Bob, who applies $X^l$ to
$b$. This is followed by a controlled-$V$ gate, \eqref{eqn9} with $b$
replacing $A$ as the control, on $b$ and $B$. Then comes a Fourier transform
\begin{equation}
F=\frac{1}{\sqrt{N}}\sum_{mj} e^{2\pi imj/N} \dyad{m}{j},
\label{eqn12}
\end{equation}
on $b$, and a measurement of $b$ in the standard basis.
The outcome $m$ is sent to Alice who carries out a $Z_m = Z^{-m}$ correction,
where $Z$ is defined in \eqref{eqn11}. [The $F$ gate could be other
than \eqref{eqn12}; see the discussion following \eqref{eqn15}.]  This
completes the protocol.

\begin{figure*}[ht]
\begin{center}
\includegraphics[scale=1]{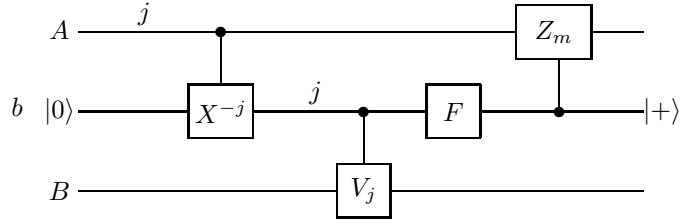}
\end{center}
\caption{Reduced diagram of the $Z$-information protocol}
\label{fgr7}
\end{figure*}

\subsection{Detailed analysis of the circuit} \label{sct3b}
To understand what is accomplished by the upper left part of the circuit in
Fig.~\ref{fgr5} it is helpful to first quantize it, see the discussion in
Sec.~\ref{sct2c}, and then show that the quantized version in
Fig.~\ref{fgr6}(a) is effectively the same thing as that in Fig.~\ref{fgr6}(b).  If
$A$ is initially in the state $\ket{j}$, unitary time development in
Fig.~\ref{fgr6}(a) involves the two steps (we omit the normalization
$1/\sqrt{N}$; note also that $l$ in the figures is equal to $k-j$)
\begin{align} &\ket{j}\otimes\sum_k\ket{k}\otimes\ket{k} \ra
\ket{j}\otimes\sum_k\ket{k-j}\otimes\ket{k} \ra \notag\\
&\ket{j}\otimes\Bigl(\sum_k\ket{k-j}\Bigr)\otimes\ket{j}
=\ket{j}\otimes\Bigl(\sum_k\ket{k}\Bigr)\otimes\ket{j}.
\label{eqn13}
\end{align} Thus at the end of these two steps we have a product state on
$\HC_{Ab}\otimes \HC_a$, with the $\HC_a$ part in the state $\ket{+} =
\sum_k\ket{k}$, \emph{independent of $j$}.  So (ii) of Theorem~\ref{thm1} is
satisfied partially: there is no information about $A$ in $a$ at the end of
the time interval shown in Fig.~\ref{fgr6}(a).  One might think of the second
controlled-$X^j$ gate as serving to ``erase'' the information about $A$ present
in $a$ at the intermediate time.  Erasing the information makes use of the
fact, which justifies the final equality in \eqref{eqn13}, that addition
modulo $N$ is a group, and for any finite group a sum over the group elements
$g$ is the same as a sum over $gh$ with $h$ fixed and $gh$ the group
product. In Sec.~\ref{sct4} below, we employ this strategy using a more general
group.

As demonstrated in \eqref{eqn13} the net result of the upper left part of the
circuit in Fig.~\ref{fgr5} is to copy the value of $j$ in $\ket{j}_A$ from
$\HC_A$ to $\HC_b$, precisely what one finds in Fig.~\ref{fgr6}(b).
This is the first step in what is sometimes called ``one bit teleportation''
\cite{1bitTeleport}, and amounts to copying a particular \emph{type} of
quantum information, the $Z$ type or its generalization to $N>2$ in the
notation of \cite{RBGtypes}, from $A$ to $b$. The nonlocal operation in
Fig.~\ref{fgr6}(b), an isometry from $A$ to the combined $A$-$b$ system, is
what is effectively accomplished by the first steps of the protocol, including
the classical communication which is essential for actually transferring the
information from Alice's laboratory to Bob's. (Note, however, that the
information is \emph{not} contained in the ``classical'' outcome of the
measurement, but resides in correlations between this outcome and system $b$;
see the analysis in \cite{RBG02}.)

When we replace the upper left part of the circuit in Fig.~\ref{fgr5} with
Fig.~\ref{fgr6}(b) and quantize the remainder, the result is Fig.~\ref{fgr7},
where the unitary time development following the controlled-$V$ gate takes the
form, assuming an initial $AB$ state $\ket{j}\otimes\ket{B}$:
\begin{align} \ket{j}\otimes\ket{j}\otimes V_j\ket{B}\ra
\sum_m\Bigl(\mted{m}{F}{j}\Bigr)\ket{j}\otimes\ket{m}\otimes V_j\ket{B}\notag\\
\ra\sum_m \Bigl(\mted{m}{F}{j}\Bigr) Z_m\ket{j}\otimes\ket{m}\otimes V_j\ket{B}
\label{eqn14}
\end{align}
For a successful unitary it is necessary, see the discussion in
Sec.~\ref{sct2c} following Theorem~\ref{thm1}, that the final state in
\eqref{eqn14} be a product of a pure state on $\HC_b$ times another on
$\HC_{AB}$.  This can be achieved if we assume that
\begin{equation} Z_m\ket{j} = c\mted{m}{F}{j}^{-1}\ket{j},
\label{eqn15}
\end{equation}
since the right hand side of \eqref{eqn14} will then be of the
desired form $c\ket{j}\otimes\Bigl(\sum_m\ket{m}\Bigr)\otimes V_j\ket{B}$.  But
for $Z_m$ in \eqref{eqn15} to be a unitary it is necessary and
sufficient that the unitary operator $F$ be such that all its matrix elements
in the standard basis are of magnitude $1/\sqrt{N}$, and then $|c|=1/\sqrt{N}$
as well.  There are many possibilities for such a unitary matrix, but one
obvious choice is the Fourier transform in Eq.~\eqref{eqn12}, and then $Z_m = Z^{-m}$, with $Z$ defined in \eqref{eqn11}.
Using this or some other appropriate choice for the $F$ and $Z_m$ gates will
result in a final state, Fig.~\ref{fgr7}, of $\ket{j}\otimes\ket{+}\otimes
V_j\ket{B}$, if the initial $AB$ state is $\ket{j}\otimes\ket{B}$.  By
linearity this extends to any other initial $AB$ state.

\subsection{Generalization to projectors $P_j$ of higher rank} \label{sct3c}
Minor and fairly obvious changes are all that is needed to extend the
discussion from rank 1 to projectors $P_j$ of general rank in \eqref{eqn9}.
First, the controlled-$X^j$ gates in Figs.~\ref{fgr5} and \ref{fgr6}, both (a)
and (b), should be replaced by the unitary operator
\begin{equation} \sum_j P_j\otimes \sum_k\dyad{k}{j+k},
\label{eqn16}
\end{equation} where the $P_j$ act on $\HC_A$, a space of any dimension
greater than or equal to $N$, and addition is mod $N$.  Second, the final
$Z_m$ gates in Fig.~\ref{fgr5} now take the form
\begin{equation}
\label{eqn17} Z_m = \sum_j \sqrt{N}\mted{m}{F}{j}^{-1} P_j = \sum_j e^{-2\pi
ijm/N} P_j.
\end{equation}
where the right side is what results when $F$ is the Fourier
transform \eqref{eqn12}.  The argument that this will produce the nonlocal
unitary $\UC$ then follows exactly the same steps as before.

\xb\outl{Discussion of Z information, dimension independence}\xa

The collection of commuting projectors $\{P_j\}$ can very well be thought of
as representing a certain type of information in the notation of
\cite{RBGtypes}, namely the type that answers the question as to which of
these properties, which $j$, is true.  Let us refer to this again as ``$Z$
information''. Then the generalized protocol functions in essentially the same
way as when the $P_j$ are of rank 1: The $Z$ information is copied onto the
$b$ system by a process that resembles one step in teleportation, here it is
used to ``choose'' which $V_j$ is to be carried out, and then erased off the
$b$ system by a process which amounts to applying a unitary correction to $A$.

A point worth making, since it will come up again, is that while the success
of the protocol, and thus the type of unitary that can be implemented this
way, depends upon precise details of the ancillary systems $a$ and $b$, it
only depends upon $A$ and $B$ in a more general way: operators on the latter
pair of systems must satisfy certain algebraic conditions, but are not
otherwise constrained.  In the case at hand the operators on $\HC_A$ must be a
family of commuting projectors that sum to the identity, while on $\HC_B$ they
must be unitaries.  Beyond this there are no requirements, which in particular
means that the dimensions of $\HC_A$ and $\HC_B$ can be arbitrarily large
compared to $N$.

\xb\section{Protocols Based on a Finite Group}
\label{sct4}\xa

\xb\subsection{Introduction to the group form of unitaries}
\label{sct4a}\xa

\xb\outl{$\UC = \sum_f U(f)\otimes W(f)$, projective representation
$\{U(f)\}$, factor system $\{\mu(f,g)\}$}\xa

The protocols discussed in this section are based on the use of a finite group
$G$, elements $f$, $g$, and so on, identity $e$, group multiplication indicated by
$fg$, in general not equal to $gf$.  We shall want to consider cases in which
a unitary $\UC$ on a tensor product $\HC_A\otimes\HC_B$ can be expressed in
the form
\begin{equation} \UC = \sum_{f\in G} U(f)\otimes W(f),
\label{eqn18}
\end{equation} where the unitary operators $U(f)$ on $\HC_A$ form a
finite-dimensional \emph{projective representation} of $G$ in the sense that
for all $f$ and $g$
\begin{equation}
U(f) U(g) = \mu(f,g) U(fg).
\label{eqn19}
\end{equation}
Here the $\mu(f,g)$ are nonzero complex numbers constituting a
\emph{factor system}; in our case they are of magnitude 1 because the $U(f)$
are unitary. We shall sometimes have occasion to use \eqref{eqn19} in the
form
\begin{equation}
\label{eqn20}
U^\dagger(g) U(f) = \mu^\ast(g,g^{-1}f) U(g^{-1}f),
\end{equation}
whose validity can be checked by left multiplying both sides by
$U(g)$, the inverse of $U^\dagger(g)$, and noting that $\mu^\ast(f,g) =
1/\mu(f,g)$.  As for our purposes it represents no loss of generality, we
shall also assume that $U(e)=I_A$, and consequently
\begin{equation} \mu(e,f) = \mu(f,e) = 1,\,\,\forall f\in G.
\label{eqn21}
\end{equation} Naturally, an ordinary representation of $G$ in which
$\mu(f,g)=1$ for every $f$ and $g$ is one possibility, but we will also
consider examples in which the collection $\{U(f)\}$ constitutes a ``group up
to phases'' with nontrivial phases.


\begin{figure*}[ht]
\begin{center}
\includegraphics[scale=1]{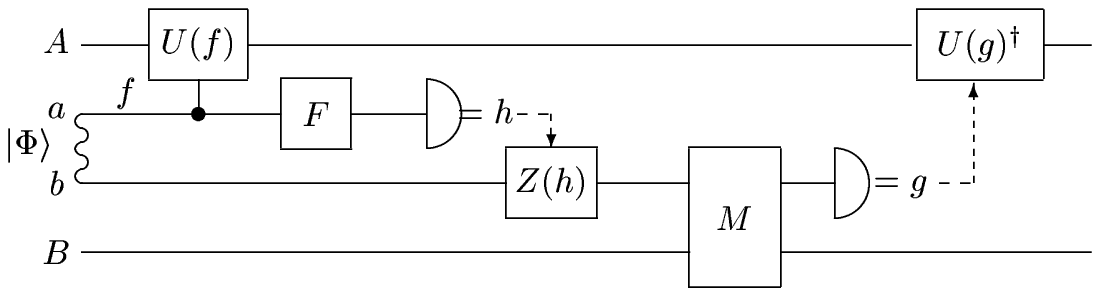}
\end{center}
\caption{Circuit diagram illustrating local implementation of nonlocal unitary
${\UC}=\sum_{f}U(f)\otimes W(f)$, where the set of unitaries $\{U(f)\}$ forms
a projective representation of a group.}
\label{fgr8}
\end{figure*}

\xb\outl{Contents of following subsections}\xa

Figure \ref{fgr8} shows a circuit that will carry out the unitary $\UC$ in
\eqref{eqn18} with the help of a suitable entangled state, as discussed in
detail in Sec.~\ref{sct4b} below, which gives sufficient conditions on $W(f)$
so the protocol will succeed.  The relationship between $\UC$ and the $W(f)$
is analyzed further in Sec.~\ref{sct4c}.  A particular case in which each
$W(f)$ is proportional to a unitary $V(f)$ representing the same group is the
topic of Sec.~\ref{sct4d}, while in Sec.~\ref{sct4e} we discuss the relationship of the expansion form \eqref{eqn18} with the controlled unitaries.

Before proceeding, let us first give a brief description of the circuit of
Fig.~\ref{fgr8}. Alice begins the protocol by implementing a controlled set of
unitaries, where system $a$ in standard basis state $|f\rangle_a$ controls the
unitary $U(f)$, which operates on $\HC_A$. She then measures $a$ in a basis
unbiased to the standard basis and defined by unitary operator $F$, where by
unbiased we mean that when represented in the standard basis, $F$ must have
all entries of magnitude $|G|^{-1/2}$. Next, Alice tells Bob the outcome $h$
of that measurement, and he follows by performing diagonal unitary $Z(h)$ on
$b$, which has the effect of removing phase factors introduced by Alice's
measurement. He then performs unitary $M$ on $bB$ (defined in
\eqref{eqn30} and \eqref{eqn31} below), which introduces the operators $W(f)$
of \eqref{eqn18} in a way that is correlated with the $U(f)$'s of Alice's
earlier operation, where these correlations are made possible by the initial
entanglement between $a$ and $b$. Bob then measures $b$ in the standard basis
and tells Alice his outcome, $g$. Alice completes the protocol with the
``correction'' $U(g)^\dagger$ on $A$, which adjusts the correlation between
$W$'s and $U$'s so that the result is always equal to $\UC$ [that is, so that
$W(f)$ is always tensored with $U(f)$, rather than with $U(f^\prime)$ for some
$f^\prime\ne f$].

\xb\subsection{Discussion of the circuit}
\label{sct4b}\xa

\xb\outl{Reduced circuit diagram }\xa

\begin{figure*}[ht]
\begin{center}
\includegraphics[scale=1]{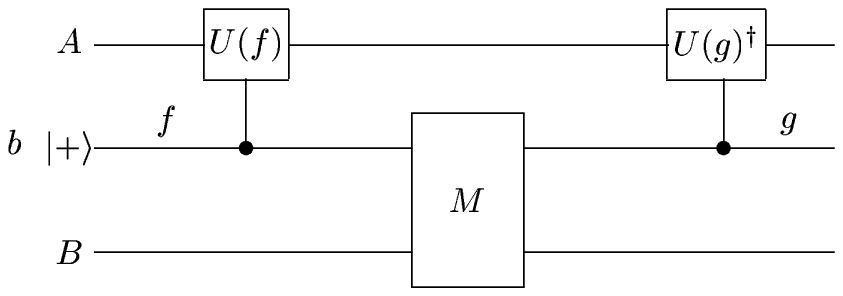}
\end{center}
\caption{Circuit equivalent to Fig.~\ref{fgr8} in terms of its action on
$\HC_A\otimes\HC_B$} \label{fig:bipartM} \label{fgr9}
\end{figure*}

Now let us analyze this protocol in detail to understand how and why it works.
In view of linearity it suffices to analyze the circuit in Fig.~\ref{fgr8} for
a product input state $\ket{A}\otimes\ket{B}$.  The resource entangled state
is
\begin{equation}
\ket{\Phi}_{ab} =\frac{1}{\sqrt{|G|}} \sum_{f\in G} \ket{f}\otimes\ket{f},
\label{eqn22}
\end{equation}
where the kets represent orthonormal basis states labeled by
elements of the group $G$ introduced previously.  The first controlled gate means
that if $a$ is in the state $\ket{f}$, $U(f)$ is carried out on $\HC_A$, while
the other controlled operations are in response to results of measurements
carried out in the standard basis. Before discussing the bipartite unitary $M$
on $\HC_b\otimes\HC_B$, it is helpful to replace the upper left part of the
circuit in Fig.~\ref{fgr8} by its counterpart in Fig.~\ref{fgr9}, in the same
way in which the upper left part of Fig.~\ref{fgr5} was simplified in
Fig.~\ref{fgr6}.  Thus after quantizing the circuit of Fig.~\ref{fgr8}, one has as in
Sec.~\ref{sct3}---compare with \eqref{eqn14}---a unitary time development
\begin{align} &\ket{A}\otimes\sum_f\ket{f}\otimes\ket{f}\ra \sum_f
U(f)\ket{A}\otimes\ket{f}\otimes\ket{f}& \notag\\ & \ra\sum_{f,h}
\mted{h}{F}{f}\; U(f)\ket{A}\otimes\ket{h}\otimes\ket{f} & \notag\\ &
\ra\sum_{f,h} \mted{h}{F}{f}\; U(f)\ket{A}\otimes\ket{h}\otimes Z(h)\ket{f}. &
\label{eqn23}
\end{align} If we choose $Z(h)$ so that
\begin{equation}
\label{eqn24} Z(h)\ket{f} = c \mted{h}{F}{f}^{-1} \ket{f},
\end{equation} the same strategy as in \eqref{eqn15}, the final term in
\eqref{eqn23} is
\begin{equation} c\sum_f U(f)\ket{A}\otimes\Bigl(\sum_h\ket{h}\Bigr)\otimes
\ket{f}, \label{eqn25}
\end{equation} thus justifying the upper left part of Fig.~\ref{fgr9}, where
the ancillary system $a$ no longer appears.  The upper right part comes from
quantizing the second measurement and classical communication in
Fig.~\ref{fgr8}, just as in Sec.~\ref{sct3}.

\xb\outl{definition of the $M$ gate, and unitary time development}\xa

To analyze the circuit in Fig.~\ref{fgr9} we begin by writing $M$ in the
block form
\begin{equation}
\label{eqn26}
M = \sum_{f,g} \dyad{g}{f}\otimes \mted{g}{M}{f},
\end{equation}
where $\mted{g}{M}{f}$ denotes an \emph{operator} on $\HC_B$, not just
a complex number.  The unitary time development represented by Fig.~\ref{fgr9}
then takes the form
\begin{align}
&\ket{A}\otimes\sum_f\ket{f}\otimes\ket{B}
\notag\\
&\ra\sum_{f,g}U^\dag(g) U(f)\ket{A}\otimes\ket{g}\otimes \mted{g}{M}{f}\ket{B}.
\label{eqn27}
\end{align} In order to have no information left in the $b$ system at the
final time---see Sec.~\ref{sct2} and compare with the analogous discussion in
Sec.~\ref{sct3}---we need
\begin{align} &\sum_f U^\dag(g) U(f)\otimes \mted{g}{M}{f}& \notag\\ &=\sum_f
\mu^\ast(g,g^{-1}f) U(g^{-1}f)\otimes \mted{g}{M}{f}&
\label{eqn28}
\end{align} to be independent of $g$, as then the right side of \eqref{eqn27}
will factor into a product of $\sum\ket{g}$ on $\HC_b$ and a pure state on
$\HC_A\otimes\HC_B$, and we will have carried out a unitary operation on the
latter.

While the $g$-independence of \eqref{eqn28} by itself does not provide a
strong constraint on the form of $\mted{g}{M}{f}$, it will be satisfied if we
set
\begin{equation}
\label{eqn29} \mted{g}{M}{f} = \mu(g,g^{-1}f) W(g^{-1}f),
\end{equation} and replace the sum over $f$ with a sum over $h=g^{-1}f$.  With
$\mted{g}{M}{f}$ in this form, \eqref{eqn26} may be rewritten as
\begin{equation}
\label{eqn30} M = \sum_{f\in G} R(f)\otimes W(f),
\end{equation} where
\begin{equation}
\label{eqn31} R(f) = \sum_{g\in G} \mu(g,f)\; \dyad{g}{gf},
\end{equation}
and the collection $\{R(f)\}$ forms a projective regular
representation of $G$ with factor system $\mu(f,g)$; see p.~267 of
\cite{GroupBook} and the comments here in Appendix~\ref{apx4}.

By using the basis $\{\ket{g}\}$ of $\HC_b$ one can view $M$ as a matrix with
blocks, where the $\mted{g}{M}{f}$ block is equal to $W(g^{-1}f)$ multiplied
by a phase $\mu(g,g^{-1}f)$. The structure is most easily visualized for a
trivial factor system $\mu(g,h)\equiv 1$ and $G$ a cyclic group, in which case
the blocks form a circulant matrix. For a more complex group structure, the
rows of $M$ are again related to each other by permuting the blocks, but now
according to the multiplication table of (the non-cyclic) $G$. Because the
$U(f)$'s represent the group $G$, this structure of $M$ makes it easy to
correct for the different measurement outcomes $g$ on $b$ by doing a unitary
correction $U(g)^\dagger$ on $\HC_A$ as shown in Fig.~\ref{fgr9}. Each of
these measurement outcomes ``picks out'' a different row of blocks in $M$, and
$U(g)^\dagger$ effectively undoes the permutation of the blocks in that row to
match each $W(f)$ with the appropriate desired $U(f)$.

The left side of \eqref{eqn28} is independent of $g$ and thus equal to its
value when $g=e$.  Inserting this on the right side of \eqref{eqn27} gives a
final state proportional to
\begin{align}
\label{eqn32}
&\sum_{f,g}U(f)\ket{A}\otimes\ket{g}\otimes\mted{e}{M}{f}\ket{B}\notag\\
&=\sum_{f}\left[U(f)\ket{A}\otimes\left(\sum_g\ket{g}\right)\otimes
W(f)\ket{B}\right],
\end{align}
which is the desired operation on $\HC_A\otimes\HC_B$. Thus in conclusion:

\begin{theorem}
\label{thm3}

If $\{U(f):\,f\in G\}$ is a projective representation of the group $G$ with
factor system $\mu(f,g)$, and the $W(f)$ are such that $M$ defined
in \eqref{eqn30} is unitary, then the circuit in Fig.~\ref{fgr8}
deterministically implements the unitary transformation $\UC = \sum_{f\in G}
U(f)\otimes W(f)$ on $\HC_A\otimes\HC_B$.

\end{theorem}

The proof is a consequence of applying Theorem \ref{thm1} to the circuit in
Fig.~\ref{fgr9}, with $\HC_R= \HC_A\otimes\HC_B$, $\HC_S=\HC_b$, and
$\JC:\HC_A\otimes\HC_B\ra \HC_A\otimes\HC_B\otimes\HC_b$ corresponding to the
unitary operation produced by the circuit with the initial $\HC_b$ state
fixed.  Since by the preceding analysis there is no information in $\HC_b$,
\eqref{eqn6} applies, which means a unitary $\UC$ is carried out on $\HC_A\otimes\HC_B$. The form of this $\UC$ is $\sum_{f\in G}
U(f)\otimes W(f)$ according to \eqref{eqn32}.  Thus given $W(f)$ such that $M$ is unitary, the
circuit will carry out the corresponding $\UC$.  The inverse problem of finding
suitable $W(f)$ operators that give rise to a unitary $M$, for a given $\UC$, is the subject of the next subsection.

\xb\subsection{Group Fourier transform}\label{sct4c}\xa

To analyze unitaries $\UC$ of the group form \eqref{eqn18} it is useful to employ
the theory of irreducible representations, which is almost the same for
projective representations as for ordinary representations with the trivial
factor system $\mu(f,g)=1$; see Ch.~12 of \cite{GroupBook} for an accessible
treatment. (The discussions there and in the following both assume that the underlying field for the vector spaces of concern is the complex number field $\mathbb C$.)

For a given group with a given factor system, there are a finite number $\kappa$ of inequivalent unitary irreducible representations $\{D^{(\lambda)}(f)\}$ labeled by an integer $\lambda$ taking values from $1$ to $\kappa$, where $D^{(\lambda)}(f)$ are $d_\lambda \times d_\lambda$ unitary matrices, which we shall assume fixed throughout the following discussion, with $\sum_{\lambda=1}^{\kappa} d_\lambda^2 = |G| = N$. By choosing a suitable orthonormal basis (independent of $f$) of $\HC_A$, the representation $\{U(f)\}$ can be written in the block
diagonal form
\begin{equation}
\label{eqn33}
U(f) = \bigoplus_{\lambda = 1}^\kappa \bigoplus_{\eta =
1}^{n_\lambda}\, D^{(\lambda)}(f) = \sum_l P_l\, U(f)\, P_l,
\end{equation}
where the irreducible representation $\lambda$ occurs with multiplicity
$n_\lambda$, and is absent from the sum when $n_\lambda=0$.  The projectors
$P_l$ that sum to the identity $I_A$ provide an alternative way of
representing the blocks, each of which corresponds to a distinct value of $l$ defined as the pair $(\lambda,\eta)$, where $\eta$ runs from $1$ to $n_\lambda$.

Let us start with a special case $\bar U(f)$ of \eqref{eqn33} in which
$n_\lambda=1$ for every $\lambda$, that is, the representation $\{\bar U(f)\}$
contains each inequivalent irreducible representation exactly once. To represent the block diagonal
structure we use an orthonormal basis in which each ket $\ket{\lm j}$ carries
two labels: $\lm$ for the representation, and $j$ an integer between 1 and
$d_\lm$, so that the matrix $\bar U(f)$ has the form
\begin{equation}
\label{eqn34}
\mted{\lm j}{\bar U(f)}{\lm' k} = \dl^{}_{\lm \lm'}D^{(\lm)}_{jk}(f)
= \dl^{}_{\lm \lm'}\tilde D(K,f).
\end{equation}
Here $\tilde D(K,f)$ is simply $D^{(\lm)}_{jk}(f)$ thought of as a matrix in
which the rows are labeled by $K=(\lm,j,k)$ and the columns by the elements
$f$ of the group $G$. It is a nonsingular $|G|\times |G|$ matrix (see further
comments in Appendix~\ref{apx3}), and consequently its $|G|$ row vectors are linearly independent, and its $|G|$ column vectors are
linearly independent.  Since each column corresponds to one of the $\bar U(f)$
matrices, we conclude that the number of linearly independent operators in the
collection $\{\bar U(f)\}$, the dimension of the linear space of operators on
$\HC_A$ that they span, is equal to $|G|$.  In effect, column $f$ of $\tilde D$
is a ``squashed'' form of the matrix $\bar U(f)$, with the elements
$\lm\neq\lm'$, which are in any case 0, omitted.

Now suppose the representation $\{U(f)\}$ contains only some and not all of
the irreducible representations.  Then \eqref{eqn34} holds with $\bar U(f)$
replaced with $U(f)$, but now some of the $\lm$'s are absent, so the rank of
the remaining part of the $\tilde D$ matrix, the number of remaining rows, is
equal to $\sum_\lm d_\lm^2$ for the remaining irreducible representations
present in the collection $\{U(f)\}$.  If, on the other hand, some of the
$n_\lm$ are larger than 1, which is to say $\{U(f)\}$ contains some
irreducible representation more than once, this corresponds to duplicating
some of the rows of $\tilde D$, which of course cannot increase its rank.
Hence we arrive at the following:

\begin{theorem}
\label{thm4}
The number of linearly independent operators in the collection $\{U(f)\}$
forming a projective representation of a group $G$ is equal to the sum of the
squares of the dimensions of the distinct inequivalent irreducible
representations contained in $\{U(f)\}$ [i.e. $\sum_\lm d_\lm^2$ for those
$\lm$ in \eqref{eqn33} for which $n_\lm$ is 1 or more].  In particular, the
$\{U(f)\}$ are linearly independent if and only if all the inequivalent
irreducible representations are present: $n_\lm \geq 1$ for every $\lm$
between 1 and $\kappa$.
\end{theorem}

Of course the block diagonal structure \eqref{eqn33} of the $\{U(f)\}$ induces
a corresponding structure in $\UC$, and again it is helpful to begin with the
case in which every inequivalent irreducible representation is present exactly
once,
\begin{equation}
\label{eqn35}
\bar \UC = \sum_f \bar U(f)\otimes W(f) = \bigoplus_{\lambda = 1}^\kappa \QCL,
\end{equation}
with each block
\begin{equation}
\label{eqn36}
\QCL = \sum_f D^{(\lambda)}(f)\otimes W(f),
\end{equation}
a $d_\lambda d_B \times d_\lambda d_B$ unitary matrix,
since each block of a block diagonal unitary is itself unitary. This can be
written explicitly using matrix elements (after choosing any orthonormal basis of $\HC_B$)
\begin{equation}
\label{eqn37}
\QC^{(\lm)}_{jp;kq} = \sum_f D^{(\lambda)}_{jk}(f) W^{}_{pq}(f),
\end{equation}
and also schematically in the matrix notation introduced in
\eqref{eqn34},
\begin{equation}
\label{eqn38}
\tilde \QC(K,L) = \sum_f \tilde D(K,f) \tilde W(f,L),
\end{equation}
where $K$ denotes $(\lm,j,k)$, $L$ denotes $(p,q)$, $\tilde D(K,f)$ is $D^{(\lm)}_{jk}(f)$, and $\tilde W(f,L)$ is $W_{pq}(f)$.

The connection between $\QCL$ and the $W(f)$ can be thought of as a group
Fourier transform whose inverse is, Appendix~\ref{apx3},
\begin{equation}
  \label{eqn39}
  W_{pq}(f) =\sum_{\lambda=1}^{\kappa}\frac{d_\lambda}{N}
  \sum_{j,k=1}^{d_\lambda}  \left[D^{(\lambda)}_{jk}(f)\right]^* \QC^{(\lm)}_{jp;kq}.
\end{equation}
This provides a convenient way of parametrizing the collection of operators
$W(f)$ for which $\bar\UC$ in \eqref{eqn35} is unitary: the parameters are the
elements of the $\QCL$ matrices subject to the sole restriction that each
matrix be unitary.  As $n^2$ real parameters are needed to characterize an
$n\times n$ unitary matrix, a total of $\sum_{\lambda=1}^{\kappa} d_\lambda^2
d_B^2=\vert G\vert d_B^2$ real parameters characterize all possible $\UC$
unitaries of the type under discussion once the basis of $\HC_A$ has been
specified as noted previously.

The special case $n_\lambda= 1$ just discussed provides the key to
understanding all other possibilities. Suppose some multiplicities $n_\lambda$
in \eqref{eqn33} are greater than 1.  Then both the $U(f)$ and the
corresponding
\begin{equation}
\label{eqn40}
 \UC = \bigoplus_{\lambda=1}^{\kappa}
 \bigoplus_{\eta=1}^{n_\lambda} \QCL =
\sum_l\left[P_l\otimes I_B\right] \UC \left[P_l\otimes I_B\right]
\end{equation}
will contain some identical blocks, but the number of parameters is still the
same: for each $\lambda$ one can choose only a single $\QCL$ in \eqref{eqn39}.
[Note the product structure of the projectors appearing on the right-hand side of \eqref{eqn40}, which is a consequence of the fact that the $U(f)$, which are the operators that represent $G$, act only on $\HC_A$.] On the other hand, if some of the $n_\lambda$ in
\eqref{eqn33} or \eqref{eqn40} are zero, then whereas the number of parameters
determining the $W(f)$ is the same as before, altering the matrix elements in
those $\QCL$ which are no longer used in the sum \eqref{eqn33} or
\eqref{eqn40} can have no influence on $\UC$.  To put it a different way,
when some irreducible representations are missing, the collection $U(f)$ is no longer linearly
independent, Theorem~\ref{thm4}, and hence $\UC$ does not uniquely determine
the $W(f)$ in \eqref{eqn35}. The right side of \eqref{eqn40} should serve as a
reminder that the block structure of $\UC$ is induced by a decomposition
of the identity on $\HC_A$, not $\HC_B$.

The next theorem summarizes these results and also gives a characterization
\eqref{eqn41} of the $W(f)$ that does not involve the group Fourier transform:

\begin{theorem}
  \label{thm5} Let $\{U(f)\}$ be a projective unitary representation of a
  group $G$ on $\HC_A$ with factor system $\{\mu(f,g)\}$, \eqref{eqn19};
  $\{W(f)\}$ a collection of operators on $\HC_B$; and
  $\UC=\sum_{f\in G} U(f)\otimes W(f)$, \eqref{eqn18}. Then

(a) The following three statements are equivalent:

\indent\indent (i) The operator $M$, defined in \eqref{eqn30} and shown in Fig.~\ref{fgr8}, is unitary.

\indent\indent (ii) The $W(f)$ are given by \eqref{eqn39} for some collection
$\{\QCL\}$ of unitary matrices.

\indent\indent (iii) The $W(f)$ satisfy
\begin{equation}
\label{eqn41}
\sum_f \mu^\ast(f,g) W^\dag(f) W(fg) = \delta(e,g) I_B,\;\forall g\in G,
\end{equation}
where $\delta(e,g)$ is 1 if $g$ is the group identity $e$ and 0 otherwise.

(b) Each of the three statements in (a) implies that $\UC$ is unitary.

(c) If $\UC$ is unitary and the $U(f)$ are linearly independent, then all
three statements in (a) hold.

(d) If $\UC$ is unitary and the $U(f)$ are linearly \emph{dependent}, so the
$W(f)$ are not uniquely determined by $\UC=\sum_{f\in G} U(f)\otimes W(f)$,
there exists at least one choice for the $W(f)$ such that the statements
in (a) are correct.

\end{theorem}

This theorem, whose proof is in Appendix~\ref{apx4}, implies that whenever a
unitary $\UC$ can be expanded in the group form \eqref{eqn18} for some projective
representation $\{U(f)\}$ there is always some choice of $\{W(f)\}$ that makes $M$
unitary, so that $\UC$ is realized by the circuit in Sec.~\ref{sct4b}.

\xb\outl{Corollary for constructing group-unitaries}\xa

The following corollary, whose proof follows immediately from (c), (a,iii), and (b) in Theorem~\ref{thm5}, is sometimes of use in constructing additional examples.

\begin{coro}
\label{thm6}
Let $\{U(f)\}$ and $\{U'(f)\}$ be two projective unitary representations of
the same group $G$ with the same factor system, and assume the $\{U(f)\}$ are
linearly independent. The unitarity of $\UC = \sum_f U(f)\otimes W(f)$ then
implies the unitarity of $\UC' = \sum_f U'(f)\otimes W(f)$.
\end{coro}

A comment on the relationship between $\UC$ and the matrix $\tilde\QC(K,L)$ of \eqref{eqn38}: the Schmidt rank of $\UC$ is the ordinary rank of $\tilde\QC(K,L)$ when rows corresponding to absent irreducible representations have been removed from the latter, a result that follows from the discussion in Sec.~\ref{sct2d}. If an irreducible representation is present more than once this corresponds to duplicated rows, which do not change the rank. In Sec.~\ref{sct5} we will sometimes find it useful to visualize the rows of $\tilde\QC(K,L)$ as the $d_B\times d_B$ blocks $B^{(\lm jk)}$ whose matrix elements are
\begin{equation}
\label{eqn42}
B^{(\lm jk)}_{pq} = \QC^{(\lambda)}_{jp;kq} =
\sum_f D^{(\lm)}_{jk}(f) W_{pq}(f),
\end{equation}
where the triple $\lm,j,k$ labeling the block corresponds to the first
argument $K$ in $\tilde\QC(K,L)$, see \eqref{eqn38}, and the indices $p,q$ to
the second argument $L$. Each block, $B^{(\lm j k)}$, is just a ``reshaping'' of the $K^{th}$ row of $\tilde\QC(K,L)$, so the (row) rank of the latter (and therefore the Schmidt rank of $\UC$) is equal to the number of linearly independent blocks $B^{(\lm j k)}$.

In summary, we have shown that the protocol of Sec.~\ref{sct4b} works for any
bipartite unitary $\UC$ of the group form \eqref{eqn18}, have given a complete
parametrization of such unitaries in terms of a set of unitary $\QCL$
matrices, \eqref{eqn40}, and have also given an explicit form for the
corresponding $W(f)$ matrices through \eqref{eqn39}.  Conversely, if the
$W(f)$ satisfy \eqref{eqn41}, or can be written in the form \eqref{eqn39}, the
corresponding $\UC$ will be unitary and can be carried out using our protocol.

\xb\subsection{Double unitary circuit}
\label{sct4d}\xa

\begin{figure*}[ht]
\begin{center}
\includegraphics[scale=1]{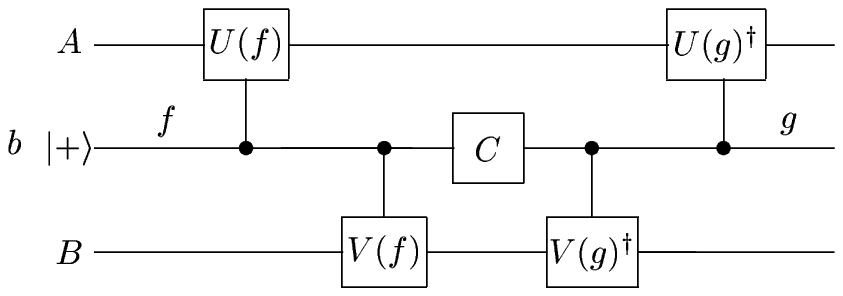}
\end{center}
\caption{``Effective'' circuit for implementing the unitary $\UC=\sum_f c(f)
U(f)\otimes V(f)$, obtained by choosing a special construction for the $M$ box
in Fig.~\ref{fgr9}.} \label{fgr10}
\end{figure*}

A particular instance of the group form \eqref{eqn18}, an extension of the example considered
in Sec.~\ref{sct2b}, is
\begin{equation}
\label{eqn43}
\UC = \sum_{f\in G} c(f) U(f)\otimes V(f) = \sum_{f\in G} c(f)\Gamma(f),
\end{equation}
where we assume the $V(f)$ are themselves a projective unitary representation
of the group $G$ with a factor system $\{\nu(f,g)\}$ analogous to the
$\{\mu(f,g)\}$ in \eqref{eqn19}.  This means that the $\Gamma(f)$ also form a
projective representation of the same group with a factor system
\begin{equation}
\label{eqn44} \gamma(f,g) = \mu(f,g)\,\nu(f,g).
\end{equation}
In the circuit that carries out $\UC$, the $M$ gate of Figs.~\ref{fgr8} and
\ref{fgr9} consists of two controlled operations on $\HC_B$ separated by a
unitary operator $C$ on $\HC_b$, as shown in Fig.~\ref{fgr10}, whose matrix
elements are
\begin{equation}
\label{eqn45} \mted{g}{C}{f} = \gamma(g,g^{-1}f) c(g^{-1}f).
\end{equation}
An equivalent definition of $C$ is
\begin{equation}
    \label{eqn46}
    C = \sum_{f\in G} c(f)R'(f),
\end{equation}
with $R'(f)$ the result of replacing $\mu(g,f)$ with $\gamma(g,f)$ in
\eqref{eqn31}.

The analysis in Sec.~\ref{sct4c} can now be applied with $\Gamma(f)$ in place
of $U(f)$, and $c(f)$ taking the place of $W(f)$. The counterparts of
Theorems~\ref{thm3} and \ref{thm5} are found in the following, the proof of which is
straightforward.

\begin{theorem}
  \label{thm7} Let $\{U(f)\otimes V(f)\}$ be a projective unitary
representation of a group $G$ with a factor system $\{\gamma(f,g)\}$,
let the irreducible representations for this factor system be
$\{D^{(\lambda)}(f):\lambda=1,\dots,\kappa\}$, and let $\UC$ be defined in
the double unitary form \eqref{eqn43}.  Then

(a) The following three statements are equivalent:

\indent\indent (i) The operator $C$ defined in \eqref{eqn45} is unitary.

\indent\indent (ii) The $c(f)$ are given by
\begin{equation}
\label{eqn47}
 c(f) = \sum_{\lambda=1}^{\kappa}\frac{d_\lambda}{N}
\sum_{j,k=1}^{d_\lambda} \left[D^{(\lambda)}_{jk}(f)\right]^*\RC^{(\lm)}_{jk}
\end{equation}
for some collection $\{\RCL\}$, where each $\RCL$ is a
$d_\lambda\times d_\lambda$ unitary matrix, with $d_\lambda$ the dimension
of the irreducible representation $\lambda$.

\indent\indent (iii) The $c(f)$ satisfy
\begin{equation}\label{eqn48}
 \sum_f \gamma^\ast(f,g) c^\ast(f) c(fg) = \delta(e,g),\;\forall g\in G,
\end{equation}
where $\delta(e,g)$ is 1 if $g$ is the group identity $e$ and 0 otherwise.

(b) Each of the three statements in (a) implies that $\UC$ is unitary and that
the circuit in Fig.~\ref{fgr8}, with the $M$ gate replaced by the $V$,
$V^\dag$ and $C$ combination as shown in the middle part of Fig.~\ref{fgr10},
will carry it out.

(c) If $\UC$ is unitary and the $U(f)\otimes V(f)$ are linearly independent,
then all three statements in (a) hold.

(d) If $\UC$ is unitary and the $U(f)\otimes V(f)$ are linearly
\emph{dependent}, so the $c(f)$ are not uniquely determined by \eqref{eqn43},
there exists at least one choice for the $c(f)$ such that the statements in
(a) are correct.
\end{theorem}

Note that $\UC$ and the $\RCL$ are related
through
\begin{equation}
\label{eqn49}
 \UC = \bigoplus_{\lambda=1}^{\kappa}
 \bigoplus_{\eta=1}^{n_\lambda} \RCL =
\sum_l \tilde P_l\;\UC\;\tilde P_l,
\end{equation}
which is similar to the relationship between $\UC$ and the $\QCL$ in
\eqref{eqn40}, except that the $\tilde P_l$ can be projectors onto entangled
subspaces, since the $\Gamma(f)$'s act on both $\HC_A$ and $\HC_B$.

\xb\outl{Understanding of matrix $C$, generating $c(f)$ by Fourier transform
from phases}\xa

To better understand what is meant by a matrix $C$ in the form \eqref{eqn45},
consider the case of a trivial factor system $\gamma(g,h)\equiv 1$. Then all
rows of the $C$ matrix are permutations of each other, generalizing the notion
of a circulant matrix (and thus called ``group circulant'' in
\cite{Klappenecker}). It can be shown that unitarity is assured if one row, say the first, is
normalized and also orthogonal to all the other rows. The same is true in the
case of a nontrivial factor system, but now the orthogonality equations
contain some additional phase factors.  Finding examples of collections $c(f)$
satisfying \eqref{eqn48}, or equivalently resulting in a unitary matrix
\eqref{eqn45} is not a trivial problem, which makes their representation
through the inverse group Fourier transform \eqref{eqn47} of some interest.
For example, if $G$ is an Abelian group with trivial factor system, the irreducible representations
are all one-dimensional phase factors and the $c(f)$ are transforms of these
phases; for a cyclic group this is just the usual Fourier transform.  Some
examples are given in Sec.~\ref{sct5}.

The Schmidt rank of $\UC$ in the double unitary form \eqref{eqn43} obviously cannot exceed the number
of linearly independent $U(f)$ or the number of linearly independent $V(f)$;
in each case this is determined, Theorem~\ref{thm4}, by which irreducible representations are
present in the representation.  In addition, the Schmidt rank cannot be larger
than the number of nonzero $c(f)$.

\xb\subsection{Relationship to controlled unitaries}
\label{sct4e}\xa

\begin{figure*}[ht]
\begin{center}
\includegraphics[scale=1]{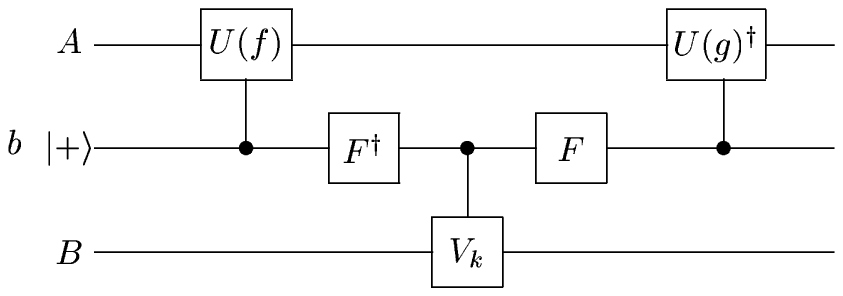}
\end{center}
\caption{Reduced diagram for implementing controlled unitaries using the
protocol in Sec.~\ref{sct4b}.} \label{fgr11}
\end{figure*}

A bipartite controlled unitary \eqref{eqn9} can always be converted to a group
unitary by defining
\begin{equation}
\label{eqn50}
U(j) = \sum_{k=0}^{N-1} \omega^{jk} P_k,\quad \omega :=\exp[2\pi i/N],
\end{equation}
with $U(0)=I$ and $U(j)U(j') = U(j+j')$, addition modulo $N$, thus a
representation of the cyclic group of order $N$.  By inverting the transform
in \eqref{eqn50} the $P_k$ can be written as linear combinations of the
$U(j)$, allowing \eqref{eqn9} to be rewritten in the form
\eqref{eqn18}.  Since in this case $G$ is
cyclic of order $N$, it is evident that the two protocols require the same
entanglement and communication resources.  Figure~\ref{fgr11} is the
counterpart of Fig.~\ref{fgr9} for this case: the $M$ gate consists of a
Fourier transform and its inverse following and preceding a controlled-$V$
gate.

Since the projectors $P_k$ in \eqref{eqn50} commute with each other, the same
is true of the $U(j)$, and this suggests that if the $U(f)$ in a
group-unitary protocol \eqref{eqn18} commute with each other it might be
possible to realize the same $\UC$ using a controlled-unitary protocol.  This
is indeed the case, for if the $U(f)$ commute with each other one can choose
an orthonormal basis of $\HC_A$ in which they are simultaneously diagonal.
This means that only one-dimensional irreducible representations occur in the
expansion \eqref{eqn33}.  Noting that there may be repeated (but equivalent) irreducible
representations, it then becomes obvious that one can write
\begin{equation}
\label{eqn51}
U(f) = \sum_\lm e^{i\phi(\lm,f)}P_\lm,
\end{equation}
where $P_\lm$ projects onto the subspace corresponding to the $\lm^{th}$
irreducible representation in the notation of \eqref{eqn33}, and
with real phases $\phi(\lm,f)$.  Upon
inserting \eqref{eqn51} into the group-unitary \eqref{eqn18} one arrives at
the controlled form \eqref{eqn9} of $\UC$ with the sum over $j$ replaced by a
sum over $\lm$.  Since the number of its one-dimensional inequivalent
irreducible (projective) representations cannot exceed the order of a group,
it is then clear that the controlled-unitary protocol requires no greater
resources than the original group-unitary protocol, and possibly less.  But if
it requires less it can always be converted back into a different, more
efficient group-unitary protocol by the procedure indicated previously.

In summary, group unitaries in which the $U(f)$ in \eqref{eqn18} commute with
each other can be converted into controlled unitaries and vice versa, with the
same requirement of resources to carry out the protocol.  If the $U(f)$
do not commute then the group-unitary protocol must be used, though it is
conceivable that $\UC$ might be realized using the controlled-unitary protocol
in a completely different way.

\xb\section{Examples}
\label{sct5}\xa

\subsection{Unitaries based on operator products $X^p Z^q$}
\label{sct5a}

If two unitary operators $X$ and $Z$ satisfy the algebraic conditions
\begin{equation}
\label{eqn52} X^n=Z^n=I, \; X Z =\omega Z X, \; \omega=e^{2\pi i/n},
\end{equation}
for some integer $n\geq 2$, the $n^2$ unitaries
\begin{equation}
\label{eqn53} U(p,q) = X^p Z^q, \quad 0\leq p,q\leq n-1,
\end{equation}
form a projective representation,
\begin{equation}
\label{eqn54} U(p,q) U(p',q') = \omega^{-qp'} U(p+p',q+q'),
\end{equation}
of the Abelian group $G$ of order $n^2$ formed by component-wise addition mod
$n$ of integer pairs $(p,q)$, where addition on the right side of \eqref{eqn54}
is to be understood as mod $n$.  In other words, $G$ is the direct product of
two cyclic groups of order $n$.

If $X$ and $Z$ in \eqref{eqn52} are the operators earlier defined in
\eqref{eqn11}, and $n=d_A$ the dimension of $\HC_A$, the unitaries
in \eqref{eqn53} form a basis of the space of operators on $\HC_A$, and
thus any bipartite unitary on $\HC_A\otimes\HC_B$ can be written in the form
\begin{equation}
\label{eqn55}
\UC = \sum_{p,q} U(p,q)\otimes W(p,q),
\end{equation}
with the expansion coefficients $W(p,q)$ operators on $\HC_B$.  As
this is of the group form \eqref{eqn18} with linearly independent $U(p,q)$ it
follows from Theorem~\ref{thm5} that it can be implemented by our protocol,
i.e., a circuit of the form shown in Fig.~\ref{fgr8} using a maximally
entangled resource state of Schmidt rank $d_A^2$.  This is precisely the same
entanglement cost as needed to teleport the $A$ state to $B$ and back again,
so there is no advantage over teleportation. However, it shows that the
protocol of Sec.~\ref{sct4} for nonlocal unitaries is as general as
teleportation.

If, on the other hand, $\UC$ can be written in the form \eqref{eqn55} for some
$n$ smaller than either $d_A$ or $d_B$, it can be carried out by our protocol
using less entanglement than teleportation. A simple example of the double
unitary type, \eqref{eqn43},
\begin{equation}
\label{eqn56}
 \UC = \sum_{p,q} c(p,q) U(p,q)\otimes U(p,q) = \sum_{p,q} c(p,q) \Gamma(p,q),
\end{equation}
takes the explicit form
\begin{align}
\label{eqn57}
\UC = &c(0,0) I\otimes I + c(1,0) X\otimes X  + \notag\\
      &c(0,1) Z\otimes Z + c(1,1) XZ\otimes XZ
\end{align}
when $n=2$.  Here the $\Gamma(p,q)$ commute with each other and form an
ordinary representation of the four group $D_2$. The $c(p,q)$ are given by
\eqref{eqn47}, where the $\RCL$ corresponding to the four irreducible
representations are complex numbers of magnitude 1. Thus the collection of all
possible coefficients can be parametrized in the form
\begin{equation}
\label{eqn58}
c(p,q) = [e^{i\alpha} +(-1)^p e^{i\beta} +(-1)^q e^{i\gamma}
+(-1)^{p+q} e^{i\delta}]/4.
\end{equation}
(It may be noted that in the case $d_A=d_B=2$, \eqref{eqn57} is the most
general two-qubit unitary up to local unitaries; see \cite{Nielsen}, which in
its Eq.~(4.3) provides an alternative representation of the $c(p,q)$
coefficients.)
For $n\geq 3$ the $\Gamma(p,q)$ in \eqref{eqn56} no longer commute with each
other, so they form a projective representation of $G$. The coefficients
$c(p,q)$ can again be parametrized using \eqref{eqn47}, with appropriate
irreducible representations, in terms of unitary matrices $\RCL$.

\subsection{Unitaries based on $S_3$ ($D_3$) }
\label{sct5b}

The nonabelian group $S_3$ of order 6, permutations of three objects, is
isomorphic to the dihedral group $D_3$.  We need only consider ordinary
representations of this group, since all factor systems are equivalent to the
trivial case $\mu(f,g)=1$; see Sec.~12.2 of \cite{GroupBook}.  There are three
inequivalent irreducible representations: $\lm=1$ (identity) and $\lm=2$ are
one-dimensional, and $\lm=3$ is two-dimensional. We omit the specific form of
the representation matrices as that is not needed in the discussion below.

The simplest interesting application to nonlocal unitaries occurs for
two qutrits, $d_A=d_B=3$, with
\begin{equation}
\label{eqn59} U(f)=D^{(2)}(f)\oplus D^{(3)}(f).
\end{equation}
One could replace $D^{(2)}$ in this formula with $D^{(1)}$ without altering
anything significant in the following discussion.  Any $\UC$ which can be
written as the direct sum of a $3\times 3$ and a $6 \times 6$ block, using an
appropriate decomposition of $I_A$ in terms of the $P_l$ introduced in
\eqref{eqn40}, can be carried out by our protocol using a maximally entangled
state of Schmidt rank 6, entanglement $\log_2 6$ (e-bits), so at a lower entanglement
cost than the $\log_2 9$ needed for two-way teleportation.  As there are five
linearly independent $U(f)$, Theorem~\ref{thm4}, the maximum Schmidt rank of
such a unitary is 5, a conclusion that also follows from the fact there are
five distinct blocks $B^{(\lm jk)}$ of the form \eqref{eqn42}. This will be
the typical Schmidt rank if the two blocks $\QC^{(2)},~\QC^{(3)}$ making up
$\UC$ are random unitaries.  Quite simple examples with Schmidt rank 5 are
easily constructed.  Here is one where the $3\times 3$ blocks of \eqref{eqn42}
that make up $\UC$ contain mostly zeros, so can be conveniently written using
dyads:
\begin{align}
\label{eqn60}
B^{(211)} &= I,\,\,\, B^{(311)} = \dyad{1}{1}+ \dyad{2}{2},\,\,\,
B^{(312)} = \dyad{3}{2},
\notag\\
B^{(321)} &= \dyad{2}{3},\quad
B^{(322)} = \dyad{1}{1}+ \dyad{3}{3}.
\end{align}

One can also construct unitaries of the double unitary form \eqref{eqn43}, for
example, with identical representations on two systems of equal dimension:
\begin{equation}
\label{eqn61}
\UC = \sum_f c(f)\, U(f)\otimes U(f) = \sum_f c(f)\Gamma(f).
\end{equation}
Since the $\Gamma(f)$ and $U(f)$ form representations of the same group, sets of
possible coefficients $c(f)$, some of them shown in Table~\ref{tbl1}, can be
generated using formula \eqref{eqn47}, with the $\RCL$ on the right side of
this equation arbitrary unitaries that serve to parametrize the possible
$c(f)$. The Schmidt rank
cannot, of course, exceed 5, and this can be achieved both with six and with
five of the $c(f)$ nonzero.  As any five of the $U(f)$ are linearly
independent, the Schmidt rank is equal to the number of nonzero $c(f)$'s when
the latter is five or less.

\begin{table}[h]
$$\begin{array}{| c c c c c c | c | c |}
\hline
  e & (123) & (132) & (12) & (23)  &  (13) & \text{SR}_3 & \text{SR}_4\\
\hline
 2/3& -1/3 & -1/3  & -1/3 &-1/3              & -1/3            & 5 & 6\\
    2/3& 1/6  & 1/6  & -i/\sqrt{3} & i/2\sqrt{3} & i/2\sqrt{3} & 5 & 6\\
    1/3& 1/3  & 1/3  & 1/\sqrt{3}  & -1/\sqrt{3}& 0            & 5 & 5\\
    1/6& -1/3 & -1/3 & -i \sqrt{3}/2 & 0          & 0          & 4 & 4\\
\hline
\end{array}$$
\caption{
The rows are four different sets of coefficients $c(f)$, for the
group elements $f$ of $S_3$ on the top line, that yield a unitary $\UC$ in
\eqref{eqn61}. The last two columns give the Schmidt ranks SR$_3$ and SR$_4$
of $\UC$ for $d_A=d_B=3$ and $d_A=d_B=4$.}
\label{tbl1}
\end{table}

Examples with $d_A=4$ where the
\begin{equation}
\label{eqn62}
U(f)=D^{(1)}(f) \oplus D^{(2)}(f)\oplus D^{(3)}(f)
\end{equation}
contain all three inequivalent irreducible representations are also easy to
construct.  In this case the $U(f)$ are linearly independent so $\UC$ may have
a Schmidt rank of 6. Whatever the size of $d_B$ the protocol only requires an
entanglement of $\log_2 6$, so it is more efficient than two-way teleportation
for $d_B\geq 3$.  Here is a collection of simple $4\times 4$ blocks, see
\eqref{eqn42}, that yields $\UC$ with a Schmidt rank of 6 when $d_B=4$.
\begin{align}
\label{eqn63}
B^{(111)} &= I,\,\,\,
B^{(211)} = \dyad{1}{1}+ \dyad{2}{2} -\dyad{3}{3} -\dyad{4}{4},
\notag\\
B^{(311)} &= \dyad{1}{1} - \dyad{2}{2},\quad
B^{(312)} = \dyad{3}{1} + \dyad{4}{2},\quad
\notag\\
B^{(321)} &= \dyad{1}{3} + \dyad{2}{4},\quad
B^{(322)} = \dyad{3}{3} - \dyad{4}{4},
\end{align}
One can also construct examples of the type \eqref{eqn61} with the $c(f)$
determined in the same manner discussed earlier, in particular the four
possibilities shown in Table~\ref{tbl1}. Now, however, the Schmidt rank is
simply the number of nonzero $c(f)$, consistent with the last column of the
table.

\subsection{Unitaries based on projective representations of $D_4$}
\label{sct5c}

The nonabelian dihedral group $D_4$ of order 8 is of interest in that in
addition to ordinary representations it also has projective representations
with a nontrivial factor system.  Corresponding to the ordinary
representations there are five inequivalent irreducible representations with
dimensions 1, 1, 1, 1, and 2.  Analyzing nonlocal unitaries based on these is
a problem similar to that discussed earlier in the case of $S_3$ ($D_3$).  The
nontrivial factor system gives rise to two inequivalent irreducible
representations, each of dimension 2, on which we now focus our attention.

The simplest situation of interest is one in which $d_A=4$ and
\begin{equation}
\label{eqn64}
U(f) = D^{(1)}(f) \oplus D^{(2)}(f),
\end{equation}
is the direct sum of the inequivalent projective representations just
mentioned.  Then using a fully entangled state of Schmidt rank 8, entanglement
of $\log_2 8$, our protocol can carry out any nonlocal unitary $\UC$ of the
form $\QC^{(1)}\oplus\QC^{(2)}$, \eqref{eqn40}, where each block $\QCL$ is a
$2d_B \times 2d_B$ matrix.

A simple example for $d_B=3$ is provided by the following collection
of block matrices, see \eqref{eqn42}:
\begin{align}
\label{eqn65}
B^{(111)} &= \dyad{1}{1} + \dyad{2}{2},\quad
B^{(112)} = \dyad{3}{1},
\notag\\
B^{(121)} &= \dyad{1}{3},\quad
B^{(122)} = \dyad{2}{2} + \dyad{3}{3},
\notag\\
B^{(211)} &= \dyad{1}{1} + \dyad{3}{2},\quad
B^{(212)} = \dyad{2}{1},
\notag\\
B^{(221)} &= \dyad{2}{3},\quad
B^{(222)} = \dyad{1}{3} + \dyad{3}{2}.
\end{align}
Another example, now for $d_B=4$, uses the $4\times 4$ block matrices:
\begin{align}
\label{eqn66}
B^{(111)}&= \dyad{1}{1} + \dyad{2}{2},\quad
B^{(112)} = \dyad{3}{1} + \dyad{4}{2},\notag\\
B^{(121)}&= \dyad{1}{3} + \dyad{2}{4},\quad
B^{(122)} = \dyad{3}{3} - \dyad{4}{4},\notag\\
B^{(211)}&= \dyad{1}{1} + \dyad{3}{3},\quad
B^{(212)} = \dyad{2}{1} + \dyad{4}{3},\notag\\
B^{(221)}&= \dyad{1}{2} + \dyad{3}{4},\quad
B^{(222)} = \dyad{2}{2} + \dyad{4}{4}.
\end{align}
The unitaries $\UC$ for both these examples have a Schmidt rank of 8.
Note that even for $d_B=3$ the entanglement cost of the protocol, $\log_2 8$, is
less than two-way teleportation of the qutrit from $B$ to $A$ and back
again.  One expects that in the generic case unitaries of this type will
have Schmidt rank 8, e.g., if they are chosen randomly subject to having the
appropriate block structure, though of course smaller Schmidt ranks can also
occur.

It is also possible to construct unitaries of the form \eqref{eqn61} using the
$U(f)$ in \eqref{eqn64}.  Again all possible sets of $c(f)$ can be generated
using formula \eqref{eqn47}, but it is important to notice that the
$\Gamma(f)=U(f)\otimes U(f)$ form a representation with a factor system
equivalent to the trivial factor system, and which can be made trivial by
choosing appropriate phases for the projective representation matrices on the
right side of \eqref{eqn64}.  Consequently, the $D^{(\lambda)}$ that appear
on the right side of \eqref{eqn47} are the five ordinary irreducible
representations of $D_4$, while the $\RCL$ in this formula are arbitrary
complex numbers of unit magnitude for $\lm=1$, 2, 3, and 4, and an arbitrary
$2\times 2$ unitary matrix for $\lm=5$.

\xb\section{Summary and Conclusions}
\label{sct6}\xa

\xb\outl{Entanglement cost}\xa

Our central result is the protocol in Sec.~\ref{sct4} which permits a unitary
$\UC$ of the group form \eqref{eqn18} to be implemented nonlocally provided the
$U(f)$ form a projective unitary representation \eqref{eqn19} of a group $G$,
and the $W(f)$ satisfy the condition \eqref{eqn41}. Since the unitarity of
$\UC$ guarantees the existence of a suitable set of $W(f)$ even when the
$U(f)$ are not linearly independent, Theorem~\ref{thm5}, the problem of
constructing an efficient scheme using our protocol comes down to minimizing
the number of terms in the sum \eqref{eqn18}, the order $N$ of the group $G$, a task accomplished in \cite{chooseGroup}.
This is because the protocol requires a uniformly entangled state of Schmidt
rank $N$ (entanglement equal to $\log_2 N$), and $2\log_2 N$ bits of classical
communication, $\log_2 N$ in each direction.

An expansion of the group form \eqref{eqn18} is possible for any bipartite unitary
$\UC$, since if $d_A$ is the dimension of $\HC_A$ it is always possible to
choose a representation of $G$ that is a basis of $d_A^2$ unitary operators for the operator space of $\HC_A$
as discussed in Sec.~\ref{sct5a}.  While our protocol works in this case, the
required entanglement and communication resources are exactly the same as
those required to teleport the quantum state of $A$ to the location of $B$ and
back again. Thus there is no gain in efficiency over simple teleportation, but
on the other hand there is also no loss in generality: our protocol can be
used for any bipartite unitary at a cost no greater than teleportation.

Thus what makes our protocol of some interest from the point of view of
efficiency is the existence of special unitaries $\UC$ for which the number of
terms in the group expansion \eqref{eqn18} is smaller than $d_A^2$, as illustrated
by the examples in Sec.~\ref{sct5}.  These special $\UC$ are, to begin with,
characterized by certain \emph{algebraic} properties of the summands in
\eqref{eqn9} or \eqref{eqn18}, properties which, in contrast to teleportation,
make no direct reference to the dimensions of either $\HC_A$ or $\HC_B$.
Hence our examples can be realized in a large number of ways on spaces of
different dimensions, and in the case of large dimensions the savings over
teleportation can be substantial.

A second useful characterization of these special unitaries is their Schmidt
rank, which is obviously a lower bound on the number of terms in
\eqref{eqn18}, thus a lower bound on the order of the group and the amount of
entanglement required for our protocol. Furthermore, Theorem~\ref{thm2}
indicates there is a similar lower bound for any other protocol based on prior
entanglement and classical communication, at least in terms of the Schmidt
rank of the entangled resource. The Schmidt rank of a nonlocal unitary on
$\HC_A\otimes\HC_B$ will typically be the minimum of $d_A^2$ and $d_B^2$, and
only if a unitary has a Schmidt rank less than this can our protocol hope to
be more efficient than teleportation.

A third requirement, not at all trivial, that must be met if a bipartite
unitary is to be efficiently realized using our protocol is that it must possess a
block diagonal form for an appropriate basis choice on one
side---$\HC_A$ in our discussion---a form commensurate with the
structure of irreducible representations of an appropriate group, in the
manner discussed in Sec.~\ref{sct4c} and illustrated in some examples in
Sec.~\ref{sct5}.  The requirement is slightly less stringent than appears at
first sight, since initial and final local unitaries might be used to
place $\UC$ in the appropriate block form.  Obviously, more study is needed in
order to give a precise characterization of the class of unitaries that can be
efficiently realized by our protocol.

\xb\outl{Use of partially entangled states}\xa

Our protocol requires a fully or uniformly entangled state, one with $N$ equal
nonzero Schmidt coefficients.  Could one instead make efficient use of a
partially entangled state of the same Schmidt rank in a deterministic protocol
of the sort we have been considering?  This would be consistent with
Theorem~\ref{thm2} if the entangling strength of the unitary were not too
great.  But even in the simplest case of a unitary on two qubits the answer is
not known (at least to us); all published protocols that use an entangled
state of Schmidt rank 2 require a fully entangled state, even if the
entangling strength of the unitary is very small.  Is there some principle of
quantum information that requires the use of a fully-entangled state?
Nondeterministic protocols or those which use an entangled state of higher
Schmidt rank are another matter; for some interesting results in this
connection see \cite{Cirac}.

\xb\outl{Considering information transfer in designing protocols}\xa

We have introduced some information theoretical considerations in
Sec.~\ref{sct2c} and subsequently seen in Secs.~\ref{sct3} and \ref{sct4} how those can
motivate or provide reasons for the choice of certain parts of the quantum
circuit, and also a picture of how information contained initially in the $A$
system can be used to influence the $B$ system without leaving an unacceptable
record in the ancillary systems.  One suspects that similar considerations might
lead to additional insights into the functioning of other efficient protocols.
Teleportation physically transports all of the $A$ information to the distant
laboratory. A more efficient protocol seems possible only if not all of this
information is really needed for the desired unitary, and the part which is
needed can somehow be separated out and transferred at less cost than full
teleportation.  Making these ideas precise and quantifying them could be
useful in designing better protocols, as well as clarifying what
quantum information is all about.

\xb\section{Acknowledgments}\xa This work has been supported in part by the
National Science Foundation through Grant Nos. PHY-0456951 and PHY-0757251. SMC
has also been supported by a grant from the Research Corporation.

\xb\begin{appendix}
\section{Proof of Theorem \ref{thm1}}\label{apx1}\xa

It is obvious that (i) implies (ii) and (iii) in Theorem~\ref{thm1}, and that
(iii) implies \eqref{eqn6} with $\WC$ times a constant in place of $\UC$.
However, as $\JC$ is assumed to be an isometry, $\WC$ times this constant must
be unitary, i.e., it preserves inner products of states $\ket{r}$ in $\HC_R$.
Thus (iii) implies (i), and to complete the proof we need only show that (ii)
implies (i) or (iii).  This follows from a quite general argument on information
location, the Somewhere Theorem of \cite{RBGtypes}, but can also be shown
directly as follows.

The absence of information from $\HC_S$ implies that if we do a measurement on
$S$ in any orthonormal basis of $\HC_S$, the probabilities of measurement
results will be independent of the input state $\ket{r}$. Consider measuring
in the $\{\ket{s_j}\}$ basis. The probability of the measurement outcome
corresponding to $\ket{s_j}$
\begin{equation}
\label{eqn67} {\rm Pr}(j)=\bra{r} K_j^{\dag} K_j \ket{r},
\end{equation}
is independent of $\ket{r}$, so $K_j^{\dag} K_j$ is proportional to the
identity $I_R$ or, equivalently, $K_j$ is proportional to a unitary operator
(since $K_j$ maps $\HC_R$ to itself).  But the same is true of any Kraus
operator $\bar K_k$ associated with a different orthonormal basis of $\HC_S$.
Thus given any two Kraus operators, say $K_1$ and $K_2$, from the original set
it follows that $(K_1+K_2)/\sqrt{2}$ and $(K_1+i K_2)/\sqrt{2}$ must be
proportional to unitaries, and from the fact that $K_1^\dag K_1^{}$, $K_2^\dag
K_2^{}$, $(K_1+K_2)^\dag (K_1+K_2)$ and $(K_1+i K_2)^\dag (K_1+i K_2)$ are
proportional to $I_R$, we conclude that $K_1^\dag K_2^{}$ is also proportional
to $I_R$, so $K_1$ and $K_2$ are both proportional to the same unitary.  By
applying this argument to every pair one sees that the $K_j$ are of the form
specified in (iii).

\xb\section{Proof of Theorem \ref{thm2}}\label{apx2}\xa

It is convenient to discuss the Schmidt rank and the entangling strength of
$\UC$ by mapping them onto properties of an entangled ket $\ket{\Omega}$ in
the following manner. Define
\begin{equation}
\label{eqn68}
\ket{\Omega_0} = \ket{\omega_\alpha}\otimes\ket{\omega_\beta}\in
\HC_{A\bar A}\otimes \HC_{B\bar B},
\end{equation}
where $\HC_{\bar A}$ and $\HC_{\bar B}$ are auxiliary systems isomorphic to
$\HC_A$ and $\HC_B$, and
\begin{equation}
\label{eqn69}
\ket{\Omega} = (\UC\otimes I_{\bar A\bar B}) \ket{\Omega_0}.
\end{equation}
If $\ket{\omega_\alpha}\propto\sum_l\ket{l}\otimes\ket{l}$ and likewise
$\ket{\omega_\beta}$ are maximally entangled states on $\HC_{A\bar A}$ and
$\HC_{B\bar B}$, the Schmidt rank of $\ket{\Omega}$ on $\HC_{A\bar A}\otimes
\HC_{B\bar B}$ is equal to that of $\UC$ by the following argument. The latter
can be defined as the number of (nonzero) terms $s$ in the Schmidt operator
decomposition
\begin{equation}
\label{eqn70}
 \UC = \sum_{j=1}^s A_j\otimes B_j,
\end{equation}
satisfying the operator orthogonality conditions
\begin{equation}
\label{eqn71}
\Tr_A(A^\dagger_j A^{}_k) = 0 = \Tr_B(B^\dagger_j B^{}_k)\text{ for $j\neq k$}.
\end{equation}
Inserting \eqref{eqn70} in \eqref{eqn69} yields
\begin{equation}
\label{eqn72}
 \ket{\Omega} = \sum_{j=1}^s (A_j\otimes I_{\bar A})\ket{\omega_\alpha}\otimes
(B_j\otimes I_{\bar B})\ket{\omega_\beta}.
\end{equation}
Given that $\ket{\omega_\alpha}$ and $\ket{\omega_\beta}$ are fully entangled
it follows that the inner products of kets of the type $(A_j\otimes I_{\bar
  A})\ket{\omega_\alpha}$ for different $j$ coincide with the operator inner
products $\Tr_A(A^\dagger_j A^{}_k)$ aside from normalization. Thus
\eqref{eqn72} is a Schmidt decomposition and the Schmidt ranks of
$\ket{\Omega}$ and $\UC$ coincide.

We are interested in simulating $\UC$ using the entangled resource
$\ket{\Phi}_{ab}$ along with separable operations (this includes LOCC).  Thus
there is a quantum operation $\{E_m\otimes F_m\}$ with $E_m:\HC_{aA} \ra
\HC_A$, $F_m: \HC_{bB} \ra \HC_B$ such that for any $\ket{\Psi}_{AB}$
\begin{equation}
\label{eqn73}
 \Bigl(E_m\otimes F_m\Bigr)\Bigl(\ket{\Phi}_{ab}\otimes\ket{\Psi}_{AB}\Bigr)
= c_m\,\UC \ket{\Psi}_{AB},
\end{equation}
where the $c_m$ are complex numbers, and the Kraus operators satisfy the usual
closure condition $\sum_m(E_m^\dagger E_m \otimes F_m^\dagger F_m) =
I_{aA}\otimes I_{bB}$. This can be extended in an obvious way to include the
auxiliary systems $\HC_{\bar A}$ and $\HC_{\bar B}$ so that
\begin{equation}
\label{eqn74}
 \Bigl(\bar E_m\otimes \bar F_m\Bigr)
\Bigl(\ket{\Phi}_{ab}\otimes\ket{\Omega_0}\Bigr)
= c_m\ket{\Omega},
\end{equation}
with $\bar E_m = E_m\otimes I_{\bar A}$ and $\bar F_m = F_m\otimes I_{\bar
  B}$. The Schmidt rank of $\ket{\Phi}\otimes \ket{\Omega_0}$ on $\HC_{aA\bar
  A}\otimes \HC_{bB\bar B}$ is the same as $\ket{\Phi}$ on $\HC_a\otimes
\HC_b$, and multiplying the former by the product operator $\bar E_m\otimes
\bar F_m$ cannot increase it. Thus the Schmidt rank of $\ket{\Omega}$ cannot
exceed that of $\ket{\Phi}$. This includes the case discussed previously in which
the Schmidt rank of $\ket{\Omega}$ is equal to that of $\UC$.  Hence the
latter cannot exceed the Schmidt rank of $\ket{\Phi}$.

The proof of the second part of the theorem uses the fact that a separable
operation applied to a pure state cannot increase the average entanglement,
see (v) in Sec.~III of \cite{Vlad_sep}, and as the entanglement of
$\ket{\Phi}\otimes \ket{\Omega_0}$ on $\HC_{aA\bar A}\otimes \HC_{bB\bar B}$,
for any choice of $\ket{\omega_\alpha}$ and $\ket{\omega_\beta}$ in
\eqref{eqn68}, is the same as $\ket{\Phi}$ on $\HC_a\otimes \HC_b$, the
entanglement of $\ket{\Omega}$ on $\HC_{A\bar A}\otimes \HC_{B\bar B}$, and
hence the entangling strength of $\UC$, cannot be greater than the entanglement
of $\ket{\Phi}$.

\xb\section{Group Fourier transform}
\label{apx3}\xa

See \cite{NielsenBook}, pp.~615, for a compact discussion of the group
Fourier transform and references to the literature. The Orthogonality Theorem
for group representations, which applies to projective as well as ordinary
representations (e.g.,~\cite{GroupBook}, p.~274), states that
\begin{equation}
\label{eqn75}
(d_\lambda/|G|)\sum_f D^{(\lambda)}_{jk}(f) D^{(\lambda')}_{j'k'}(f)^*
=\delta_{\lambda\lambda'}\delta_{jj'}\delta_{kk'},
\end{equation}
where $D^{(\lm)}_{jk}$ is the unitary matrix representing the group element
$f$ in the irreducible representation $\lm$; note that all these representations
belong to the same factor system.  Now \eqref{eqn75} is equivalent to the
assertion that $\hat D(K,f)$, with $K=(\lm,j,k)$, defined by
\begin{equation}
\label{eqn76}
\hat D(K,f) = \sqrt{d_\lm/|G|}\;\tilde D(K,f) =
\sqrt{d_\lambda/|G|}\,D^{(\lambda)}_{jk}(f),
\end{equation}
has orthogonal and normalized rows, thus is a unitary matrix. Consequently it
is nonsingular (and the same is true of $\tilde D(K,f)$), and its inverse is
the matrix $\hat D(f,K)^*$.  This allows one to invert \eqref{eqn38} after
multiplying both sides by $\sqrt{d_\lm/|G|}$, and the result is \eqref{eqn39}.

\xb\section{Proof of Theorem~\ref{thm5}}
\label{apx4}\xa

Before proving the theorem, we first prove that the $R(f)$'s defined in \eqref{eqn31} form a projective representation of $G$. This can be easily checked:
\begin{align}
\label{eqn77}
 &R(f)R(g) = \sum_{h,k} \mu(h,f)\mu(k,g) \ket{h}
\inpd{hf}{k}\bra{kg}
\notag\\
 &= \sum_h \mu(h,f)\mu(hf,g)\dyad{h}{hfg} = \mu(f,g)R(fg),
\end{align}
where the final equality is a consequence of the associative rule
\begin{equation}
\label{eqn78}
 \mu(h,f)\mu(hf,g) = \mu(h,fg)\mu(f,g).
\end{equation}
for the factor system.  [See, e.g., Ch.~12 of ~\cite{GroupBook}, where $R(f)$
is identified as the projective regular representation.]  What is important for
our purposes is that, just as in the case of ordinary representations, it
contains each inequivalent irreducible representation $\lambda$ of $G$ a
number of times equal to $d_\lambda$.

Our proof of Theorem~\ref{thm5} begins with the connection of (iii) in (a) to
(b) and to (c). From \eqref{eqn18} and \eqref{eqn20} it follows that
\begin{equation}
\label{eqn79}
{\UC}^\dag{\UC} = \sum_{f,g\in G} \mu^\ast(f,g) U(g)\otimes W(f)^\dag W(fg).
\end{equation}
Thus if the $W(f)$ satisfy \eqref{eqn41}, ${\UC}^\dag{\UC}=U(e)\otimes I_B=
I_A\otimes I_B$, and $\UC$ is unitary.  On the other hand, if the $U(f)$ are
linearly independent, for every $g$ the coefficient (an operator on $\HC_B$)
of $U(g)$ on the right side of \eqref{eqn79} is uniquely determined by the
left side, so the unitarity of $\UC$ implies \eqref{eqn41}.

The proof of (a), (b), and (c) will thus be complete once we show the
equivalence of the three statements in (a). The equivalence of (i) and (iii)
follows from \eqref{eqn30} upon replacing $\UC$ in the preceding argument by
$M$ and $U(f)$ by $R(f)$, and noting that the $R(f)$ are linearly independent,
since for each $f$ the right side of \eqref{eqn31} involves a unique set of
dyads.  That (i) implies (ii) comes from noting that with $U(f)$ replaced by
$R(f)$ in the discussion in Sec.~\ref{sct4c}, $M$ can be written in the form
\eqref{eqn40}, and the unitarity of each $\QCL$ is a consequence of
the unitarity of $M$. Thus the $W(f)$ are indeed given by \eqref{eqn39} for a
collection of unitary $\QCL$.  Conversely, if \eqref{eqn39} holds for
some collection of unitary $\QCL$, then $M$ expressed in the form
\eqref{eqn40} will be unitary, so (ii) implies (i).

To prove (d), note that by Theorem \ref{thm4} if the $U(f)$ form a dependent
collection, some of the irreducible representations must be missing, and thus $\UC$ determines,
through \eqref{eqn40}, unitary $\QCL$ matrices for only some values
of $\lambda$.  But we can choose arbitrary unitary matrices for the remaining
values of $\lambda$ without changing $\UC$, and use the resulting
$\QCL$ collection to define a set of $W(f)$ operators by means of
\eqref{eqn39}.  As this set satisfies condition (ii) of part (a) of the
theorem, it satisfies (i) and (iii) as well.\hfill\qedsymbol

\xb
\end{appendix} \xa

\xb


\begin{thebibliography}{99}
\bibitem{note_framework}
See for example the discussion in Ch.~17 of \cite{ConsistentQT}.
The earliest use of this principle in a quantum information theory context
appears to be \cite{Grif_Niu}.  One consequence is the calculational rule
stated in \cite{NielsenBook}(p.~186; also Exercise 4.35), where unfortunately the intuitive basis of the rule is not stated.
\end{thebibliography}
\end{document}